%% file: MER-method-v15_CK_PP.tex
\documentclass[12 pt]{iopart}
\usepackage{graphicx}
\usepackage{dcolumn}
\usepackage{bm}
\usepackage{graphicx,pstricks,color} 
\usepackage{upgreek}
\usepackage{enumerate}
\usepackage[mathlines]{lineno}
\usepackage{listings}
\usepackage{subfigure}

\usepackage{multirow}
\usepackage{tabularx,ragged2e}

\usepackage{array}

\graphicspath{{Figs/}}

\begin{document} 

\title {Investigation of Soft and Living Matter using a Micro-Extensional Rheometer 
}

\author{Sushil Dubey$^\dagger$, Sukh Veer$^\dagger$, Seshagiri Rao R V$^\dagger$, Chirag Kalelkar$^{\#}$, and Pramod A Pullarkat$^{\dagger *}$}

\address{$^\dagger$ Soft Condensed Matter group, Raman Research Institute, C. V. Raman Avenue, Bengaluru 560080, India \\
$^{\#}$ Department of Mechanical Engineering, Indian Institute of Technology, Kharagpur 721302, India
}
\ead{pramod@rri.res.in
} 

\begin{abstract}
Rheological properties of a material often require to be probed under extensional deformation. Examples include fibrous materials such as spider-silk, high-molecular weight polymer melts, and the contractile response of living cells. Such materials have strong molecular-level anisotropies which are either inherent or are induced by an imposed extension. However, unlike shear rheology, which is well-established, techniques to perform extensional rheology are currently under development and setups are often custom-designed for the problem under study. In this article, we present a versatile device that can be used to conduct extensional deformation studies of samples at microscopic scales with simultaneous imaging. We discuss the operational features of this device and present a number of applications.  
\end{abstract}

\maketitle 

\section{Introduction}
Rheological characterisation of materials is usually performed by subjecting them to a shear deformation with feedback control on either the applied shear stress or strain \cite{barnes1989}. However, there are several situations where properties of viscoelastic materials need to be probed under extensional deformations, either due to their geometry (filamentous materials) or to structural anisotropies in the material \cite{mckinley2002}. Examples include the investigation of mechanical properties of natural silk fibers \cite{swanson2006, liu2005b}, reconstituted silk \cite{koeppel2018, kojic2006}, synthetic fibers \cite{sharma2018}, polymer melts under extensional flow \cite{mckinley2002}, and studies of living cells such as contractile fibroblasts \cite{pablo, thoumine1997} and axons of neuronal cells \cite{bernal2007, dubey2020-axon}. Biological systems are of especial interest as they form a class of nonequilibrium systems that can generate active internal stresses, driven by molecular motors \cite{pablo, bernal2007}. 

Extensional rheometry, where well-prescribed extensional strain protocols can be applied to a sample, is underdeveloped \cite{mckinley2002}. This is, to a large extent, due to challenges involved in designing suitable instruments as the instrument design often depends on the sample one needs to study. For example, investigation of polymer melts under extensional flow is typically carried out by pulling strands which are $>$1 mm in diameter and $>$1 cm in length \cite{book-polymer-melts}. However, biological samples such as living cells need to be maintained in a fluid medium and are microscopic in scale (a few tens of microns), and require to be pulled horizontally if simultaneous observations using an inverted microscope are to be made. Therefore, unlike shear rheometry, extensional rheometry demands a wider range of techniques, which are not often commercially available.

In this article, we exhibit a device that we have developed to perform the extensional rheometry of microscopic samples or samples available only in nano- to microliter volumes. Micro-scale experiments are also relevant when thin fluid strands need to be investigated, wherein surface effects can be significant. This device is ideally suited to investigate viscoelastic and contractile responses of biological materials such as living cells, as it is built around an inverted optical microscope. This device, which we call the Micro-Extensional Rheometer (MER), has been used to perform extensional rheometry using different protocols on a variety of systems, both inanimate and living. We summarise some of these applications of the device below.

\section{Micro-Extensional Rheometer (MER)}

\begin{figure}[!h]
\begin{center}
\includegraphics[width=4.6in]{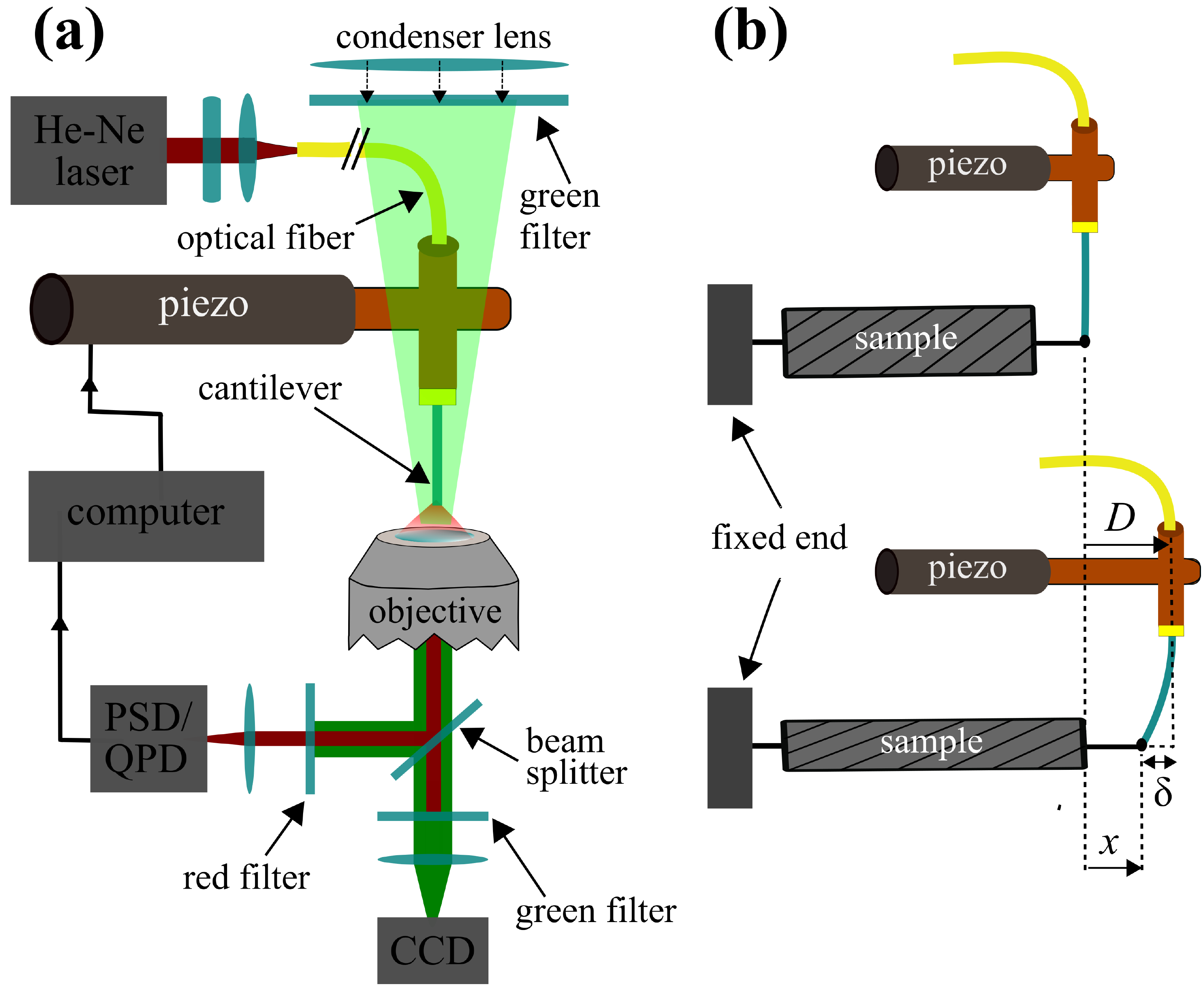}
\caption{{\bf Schematic of the Micro-Extensional Rheometer (MER).} {\bf (a)} A schematic of the MER that uses an optical fiber as a cantilever to stretch microscopic samples. Stretching is done using a closed-loop piezoelectric transducer. The tip of the cantilever is tracked by imaging the laser light exiting the cantilever onto a Position Sensitive Detector (PSD) or a Quadrant Photo-Diode (QPD). The sample is illuminated using green light which can then be separated from the red laser light using appropriate filters as shown. A computer interface links the PSD/QPD and the piezoelectric transducer for implementing closed-loop operations. {\bf (b)} A schematic representation of a viscoelastic material being stretched by translating the piezoelectric transducer by a distance $D$ resulting in a sample extension $x$. The force is computed by using the cantilever deflection $\delta$ and piezo displacement, as $F=-k(D-x)=-k\delta$, where $k$ is the cantilever force constant. 
}
\label{schematic-setup}  
\end{center}
\end{figure}
\begin{figure}[!h]
\begin{center}
\includegraphics[width=4.0in]{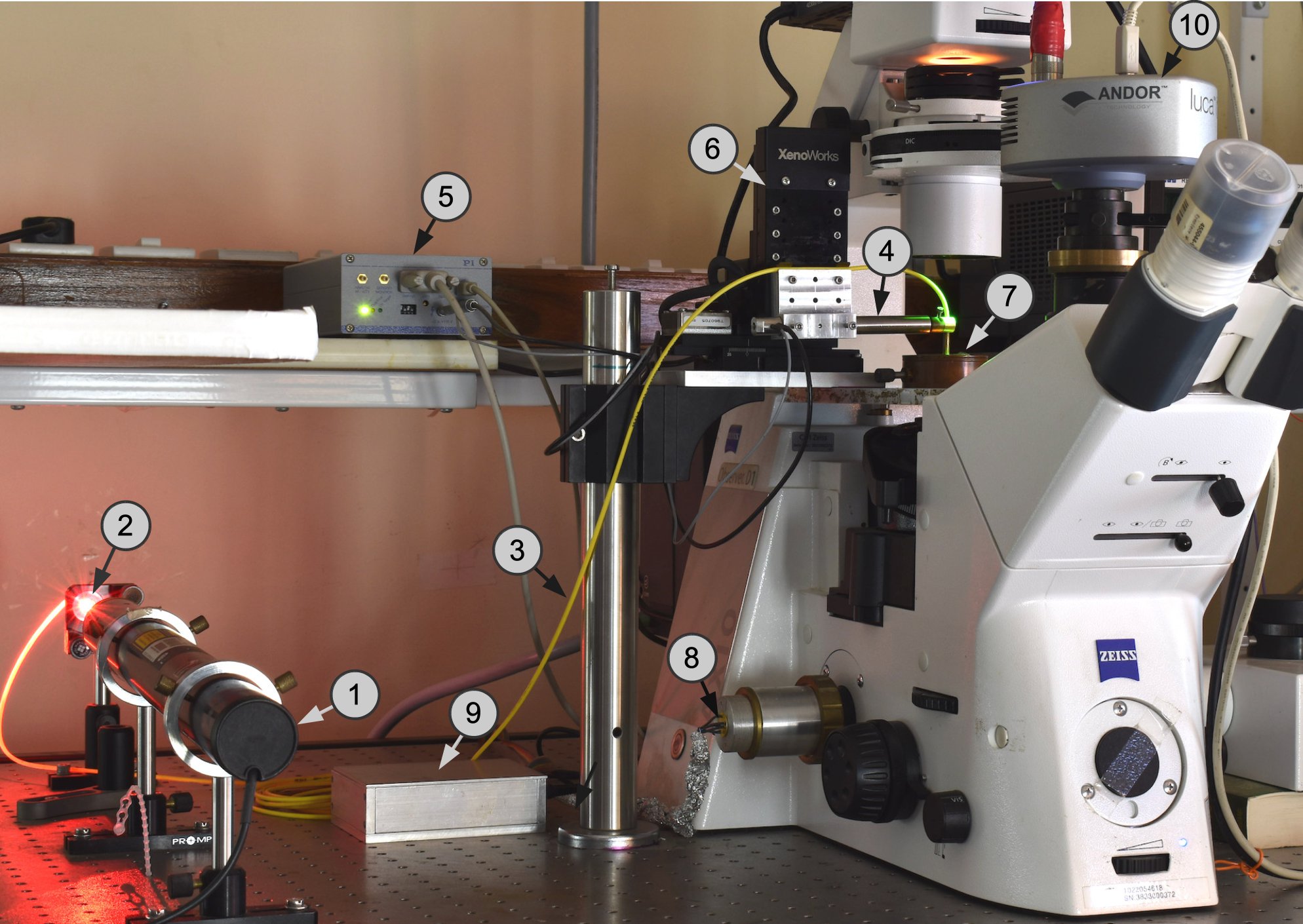}
\caption{{\bf A photograph of the MER.} Labels indicating: 1-He-Ne laser, 2-fiber coupler, 3-single mode optical fiber, 4, 5-piezoelectric transducer and controller, 6-motorized stage and joystick-controlled XYZ positioning system, 7-sample incubator, 8, 9-Position Sensitive Detector and electronics, 10-CCD camera.
}
\label{MER-photo}  
\end{center}
\end{figure}

In this section, we describe the basic operating principles of the Micro-Extensional Rheometer and direct the reader to Rao {\it et al.} \cite{giri2013} and Paul {\it et al.} \cite{paul2017} for technical specifications. The MER is based on a cylindrical cantilever fabricated by uniformly etching one end of a glass optical fiber (fiber diameter $=125\pm1$ $\mu$m, core diameter $\simeq4.3$ $\mu$m). The etching is performed by dipping the required length of the bare glass fiber, typically 5-20 mm, sequentially into solutions of hydrofluoric acid of decreasing concentrations. The acid is gently stirred using a magnetic stirrer and the time required for a certain diameter can be calibrated. After etching, the tip of the etched fiber is cut to obtain a circular exit aperture, as indicated by the symmetry of the laser light exiting the fiber. By this method, cantilevers as thin as 10 $\mu$m and as long as 20 mm can be fabricated with a taper of less than a few microns, and calibrated as detailed below. This allows control of the force, which can range between a few piconewtons to a few hundred micronewtons, depending on the cantilever dimensions and the method of detection. The base of the cantilever is then attached to a closed-loop, linear, piezoelectric transducer, with a travel range of 94 $\mu$m and a measured resolution of 10 nm, such that the cantilever hangs vertically above the objective lens of an inverted microscope (figure~\ref{schematic-setup}). The unetched end of the optical fiber is coupled to a 17 mW Helium-Neon laser (wavelength $\lambda = 632.8$ nm) using a  fiber connector and a collimator. The intensity of the laser is regulated using either a polariser or a neutral-density filter placed between the laser and the collimator. The laser light exiting the cantilever-end of the fiber is collected by the microscope objective, and the exit aperture of the fiber is imaged onto either a Position Sensitive Detector (PSD) or a Quadrant Photo Diode (QPD) mounted on one of the camera ports of the microscope (figure~\ref{schematic-setup}). The sample under study is placed in the focal plane of the objective and is stretched in this plane, while the resulting deflection of the cantilever is monitored by the PSD or QPD. This allows for real-time imaging of the deformation of the sample using a Charge-Coupled Device (CCD) camera mounted on the microscope. The imaging is done using green light obtained by placing an interference filter inside the microscope condenser, so that the green illumination light and the red laser light exiting the tip of the cantilever can be separated using appropriate filters placed in front of the detector and the camera. The detector (PSD or QPD) is read using appropriate electronics and a data-acquisition card to acquire position information with the aid of a computer \cite{giri2013}. The choice of detector is based on the application: PSD, being an area detector, can record large deformations/strains whereas the QPD can only record small displacements as the laser spot has to sufficiently illuminate all four quadrants, but gives a higher spatio-temporal resolution. Note that the deflection of the tip of the cantilever is magnified by the microscope, providing an optical amplification factor which can be selected using suitable objectives. With a 40$\times$ objective, the PSD can measure displacements up to $\sim$90 $\mu$m with a resolution of 35 nm whereas the QPD can measure a maximum displacement of  $\sim$1 $\mu$m  with a resolution of  $\sim$4 nm. The force on the cantilever is computed from the displacement imposed by the piezoelectric transducer $D$ and the shift in the cantilever tip position $x$ as $F = -k(D-x)$, where $k$ is the force constant for the cantilever (see figure~\ref{schematic-setup}). A feedback loop is used for controlling strain and it works in a proportional mode wherein the proportionality constant is optimized for different types of experiments based on the cantilever bending stiffness and typical sample response. The earlier version of this device reported in Rao {\it et al.} \cite{giri2013}, did not have the ability to probe frequency response. This shortcoming has been overcome subsequently by including  different protocols such as oscillatory extensional strain, sequential step-strain, cyclic ramp, {\it etc.} \cite{dubey2020-axon, paul2017, dubey2019-silk}. If a PSD or QPD is unavailable, the cantilever tip deflection can be measured using a CCD camera by finding the intensity-weighted centroid of the laser spot \cite{giri2013}. While this is straightforward, implementing real-time force measurement with feedback for strain control is onerous and yields relatively lower temporal resolution. For all three detection methods, the intensity distribution of the laser light should have a single prominent maximum and remain unaltered and unobstructed during an experiment. The various extensional strain protocols implemented using this MER setup were extensively tested using a mock sample with known characteristics and the data is shown in Supplementary Material figures S1 to S4. A photograph of the MER setup is shown in figure~\ref{MER-photo}.

\subsection{\bf Calibration of the optical fiber}

\begin{figure}[!h]
\begin{center}
\includegraphics[width=3.0in]{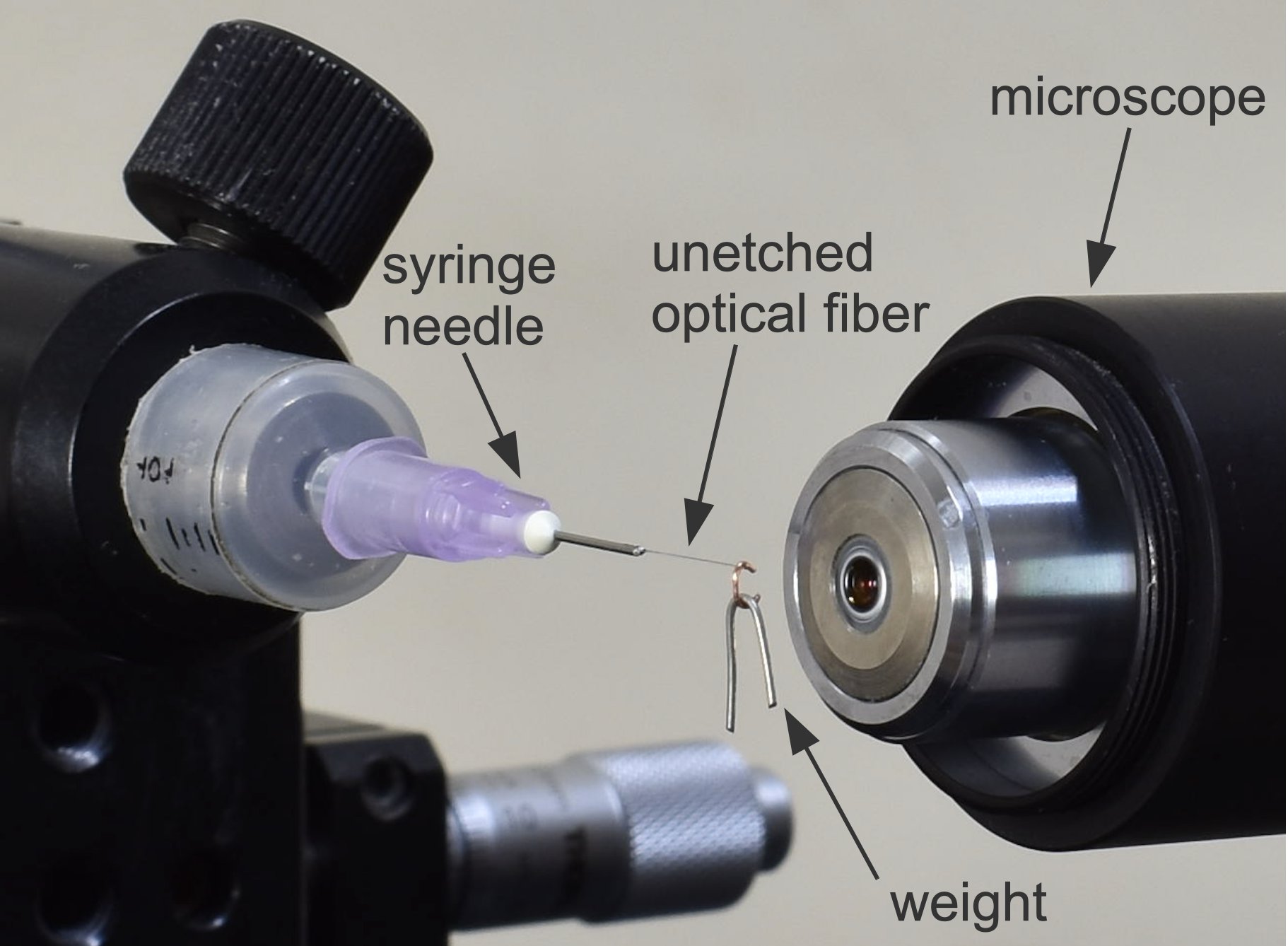}
\caption{ {\bf Cantilever calibration setup.} A piece of the unetched glass optical fiber is inserted in a syringe needle with the pointed end cut to obtain a circular opening, which acts as a holder for the fiber. A tiny amount of superglue is applied on the end of the needle before inserting the fiber to tightly clamp the fiber at its base. The needle is attached to a XYZ micrometer stage to adjust its position for imaging. Thin metal wires are loaded onto the free end of the cantilever. A 20$\times$ objective, a tube lens, and a camera are used to record the images, to calculate the deflection of the cantilever.
}
\label{calibration-setup}  
\end{center}
\end{figure}
The force constant of the cantilever is obtained by placing the cantilever in a horizontal position and loading it with small pieces of thin metal wires of known weights. The deflection is measured using a microscope placed horizontally and a Complementary Metal-Oxide Semiconductor (CMOS) camera (DCC1545M-GL, Thorlabs Inc., United States), as shown in figure~\ref{calibration-setup}. For a cylindrical cantilever with a load suspended at the tip, the Young's modulus of the material is given by $E=(FL^3)/(3I\delta)$ \cite{landau}; where $L$ is the length of the cantilever, $F$ is the force exerted by the load, $I$ is the area moment of inertia of the cantilever and $\delta$ is the deflection of the cantilever. The force constant $k$ can be obtained by using $k=(3\pi ER^4)/(4L^3 )$ \cite{landau}; where $R$ is the radius of cantilever.

\begin{figure}[h!]
\begin{center}
\includegraphics[width=4in]{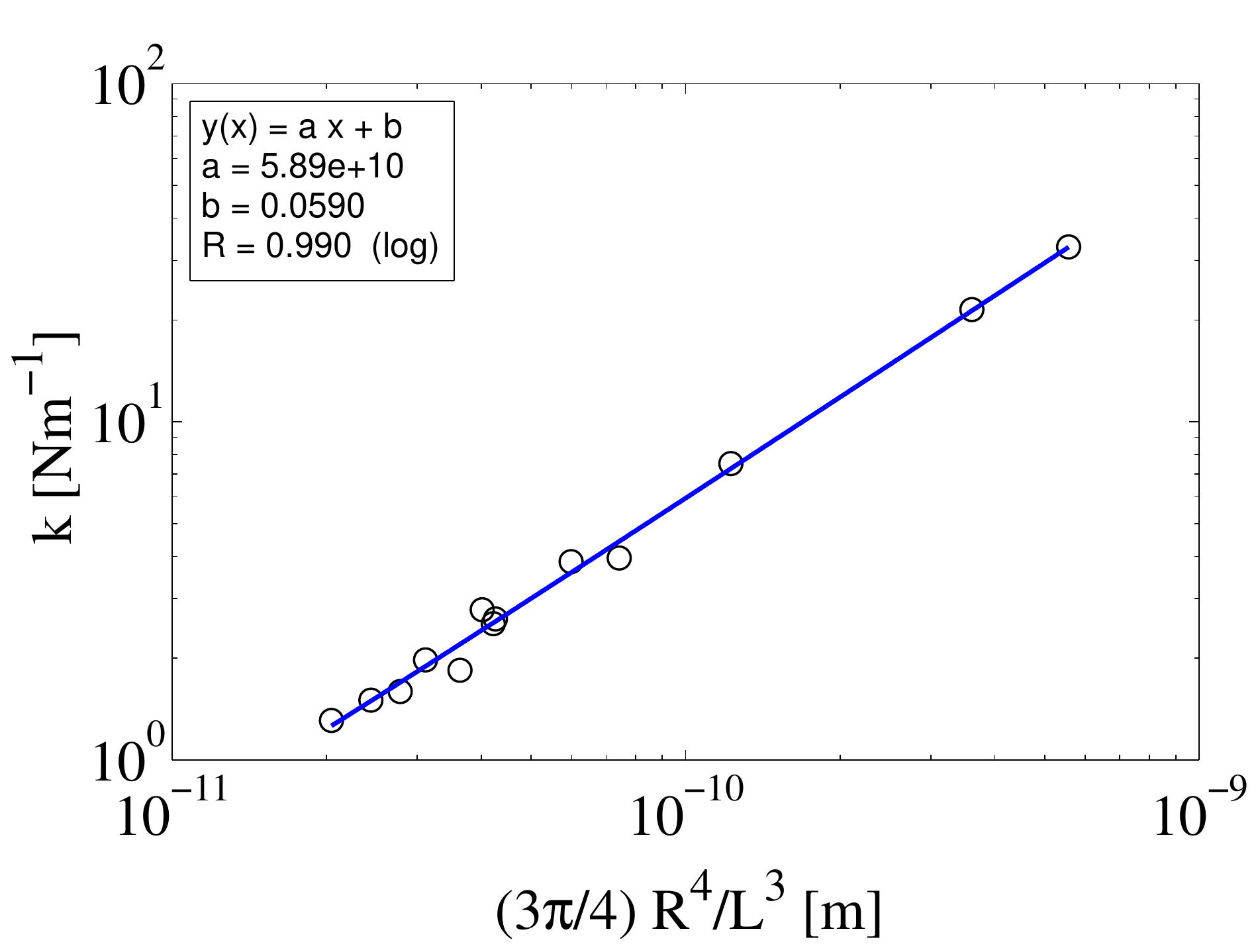}
\caption{\textbf{Cantilever calibration plot.} A log-log plot showing the dependence of the cantilever force constant on its radius $R$ and length $L$. The data, obtained by using unetched fibers of different lengths loaded with different known weights, is fitted to a straight line, the slope of which gives the Young's modulus of the glass cantilever. Each data point is averaged over at least ten measurements using different weights for the same cantilever.
}
\label{cantilever-calibration}  
\end{center}
\end{figure}

For a fixed length (unetched) cantilever, we have used different weights and calculated the value of $k$ from each loading, to obtain an average value of $k$. We repeated this procedure for different  lengths of cantilevers to obtain the plot shown in figure~\ref{cantilever-calibration}. The slope of the plot gives the Young's modulus of the cantilever which is found to be 58.9 GPa.

\subsection{{\bf Fluorescence imaging}}

\begin{figure}[h!]
\begin{center}
\includegraphics[width=4in]{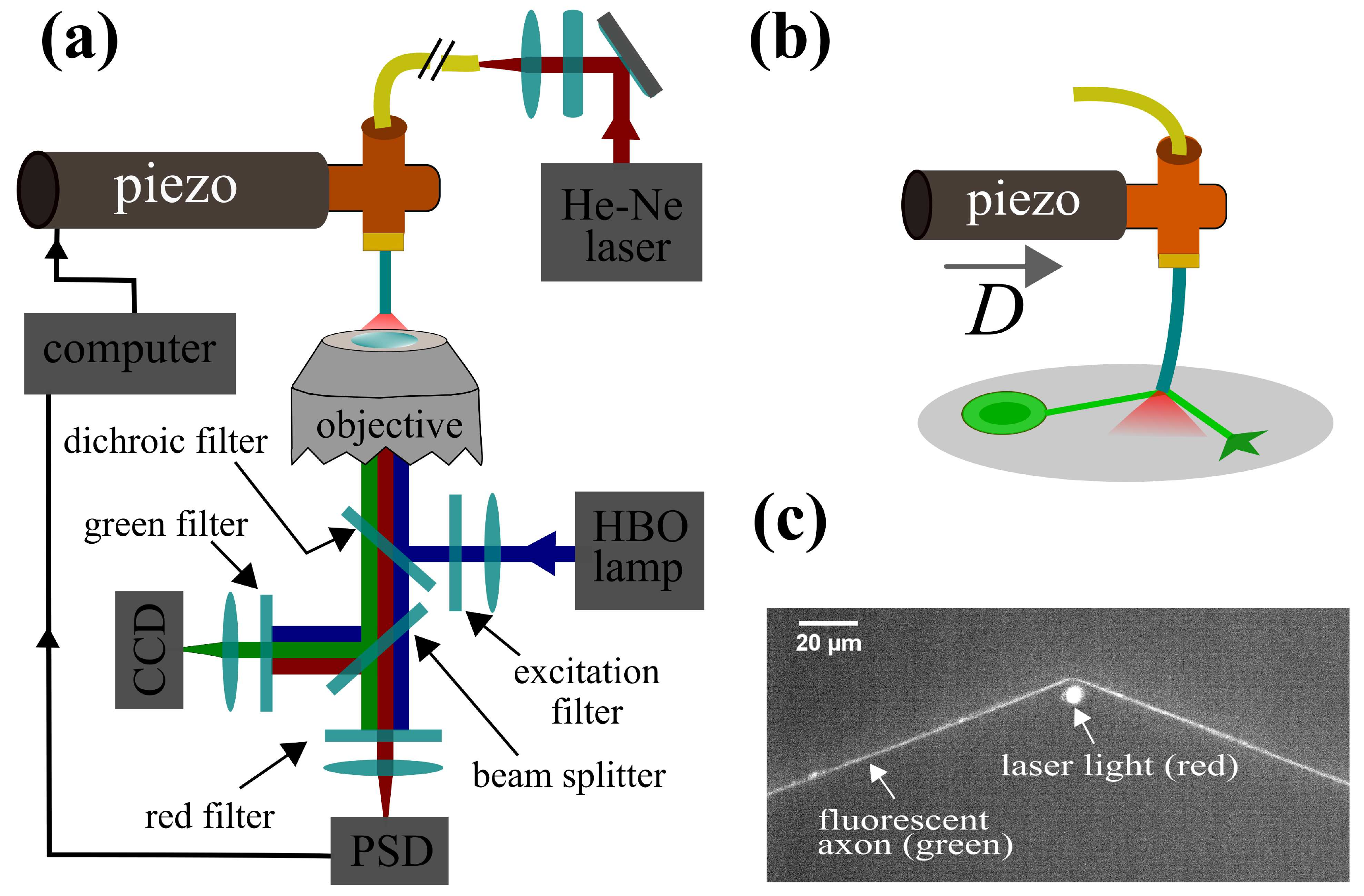}

\caption{{\bf Fluorescence imaging mode.} {\bf (a)}
Fluorescence imaging is enabled using a short-arc mercury lamp (HBO) and an epifluorescence filter unit of the Zeiss AxioObserver D1 microscope (Zeiss filter set 13: Ex. BP 470/20, dichroic FT 495, Em. 505-530). The emission filter is removed from this unit and placed in front of the CCD camera to allow the laser light exiting the cantilever tip to reach the PSD. {\bf (b)} A schematic showing how the axon of a genetically labelled neuronal cell is stretched. {\bf (c)} An image of a stretched axon which is genetically labelled with soluble Green Fluorescent Protein. The green fluorescence light and the red laser light exiting the cantilever are simultaneously imaged via a monochromatic CCD camera.
}
\label{setup-fluorescence}  
\end{center}
\end{figure}

One of the major advantages of the MER is that it permits the inspection of structural changes occurring within a sample that is subjected to extensile deformations. As opposed to standard atomic force microscopes, the sample deformation in the case of MER occurs in the focal plane of the microscope. Most optical microscopy techniques such as polarising microscopy, epifluorescence microscopy, confocal microscopy, {\it etc.} are compatible with the MER. As proof of principle, we demonstrate a scheme to implement fluorescence microscopy which is of paramount importance in investigating biological samples. For this, the emission filter is removed from the epifluorescence filter unit of the microscope and inserted in front of the monochromatic CCD camera. This allows the laser light to reach the PSD without being cut off by the emission filter as shown in figure~\ref{setup-fluorescence}(a). As a demonstration experiment, we have genetically modified a neuronal cell to express soluble Green Fluorescent Protein (GFP) in its cytoplasm and the axon (tubular extension of the neuron) is stretched using an etched optical fiber as shown schematically in figure~\ref{setup-fluorescence}(b). A fluorescence image of a stretched axon along with the laser light exiting the cantilever is shown in figure~\ref{setup-fluorescence}(c).
 
\section{Rheological measurements using MER}

In the following sections we describe several applications of the MER which has enabled us to probe polymer melts, living cells, spider silk, and bacterial suspensions in a manner which is not possible using existing techniques. The sections aim at 
describing techniques and details of interpretation of the data, where available, may be found in the citations provided in each section.

\subsection{\bf Extensional rheology of polymer melts}

\begin{figure}[h!]
\begin{center}
\includegraphics[width=6in]{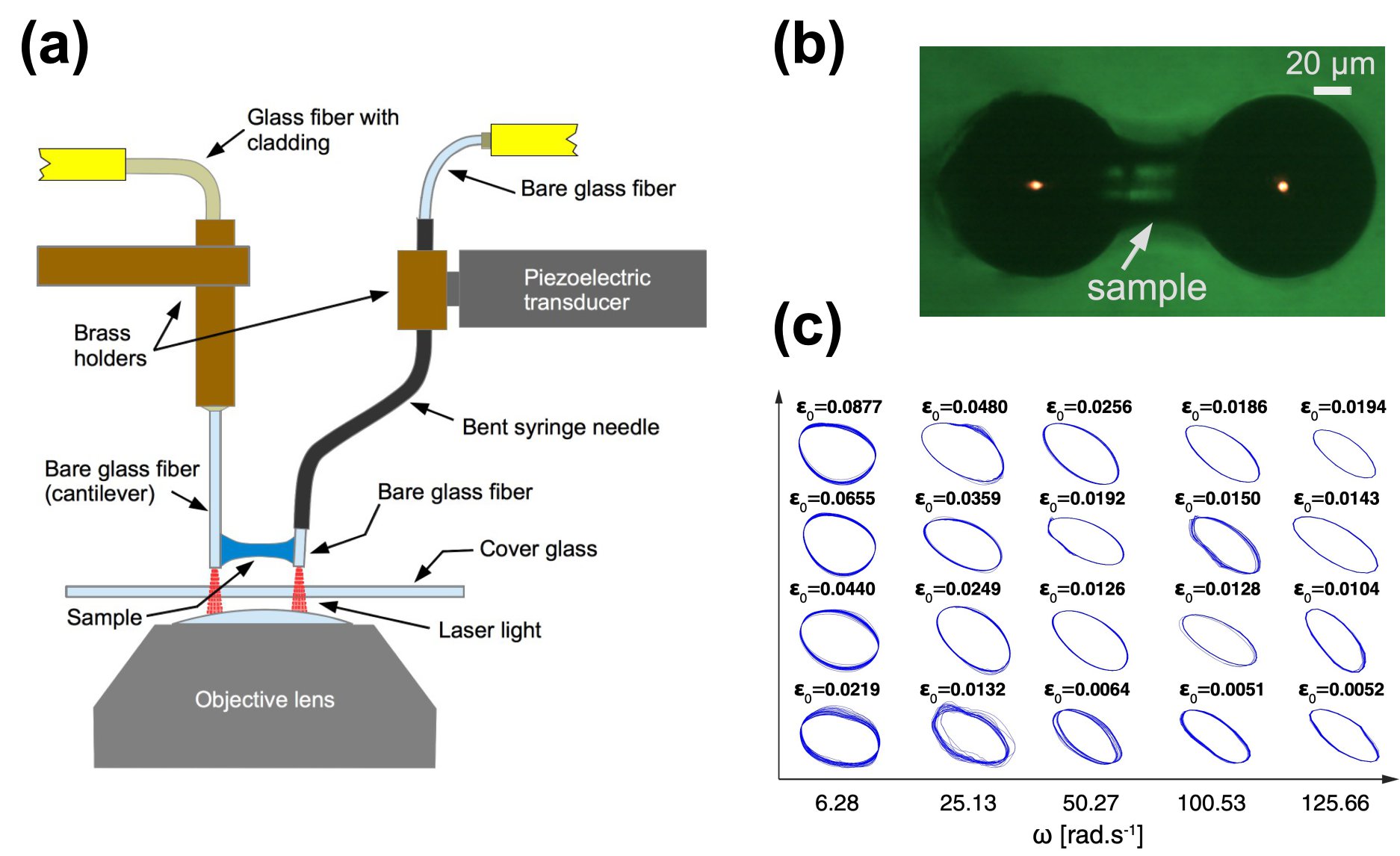}
\caption{
{\bf Oscillatory extensional strain on fluid filaments.} {\bf (a)} A schematic of the experimental setup used to investigate nanoliter volumes of fluid samples under oscillatory extensional strain. A dual optical fiber arrangement was used to track both the imposed sinusoidal oscillations on the rigid surface and the resulting response of the cantilever using a CCD camera. {\bf (b)} An image of the dual fiber arrangement with a polymer melt suspended between the two fibers as seen through the microscope. The two laser spots (red dots) are tracked at a frame rate of 400 Hz and analysed using a  centroid detection method (also see Supplementary Material Video-1). {\bf (c)} Pipkin diagrams obtained using the oscillatory extensional strain protocol for a polydimethylsiloxane (PDMS) filament. Figures 6(a) and 6(c) are reproduced with permission from Ref. \cite{paul2017}.
}
\label{exten-rheo}  
\end{center}
\end{figure}

\begin{figure}[!h]
\begin{center}
\includegraphics[width=3.5in]{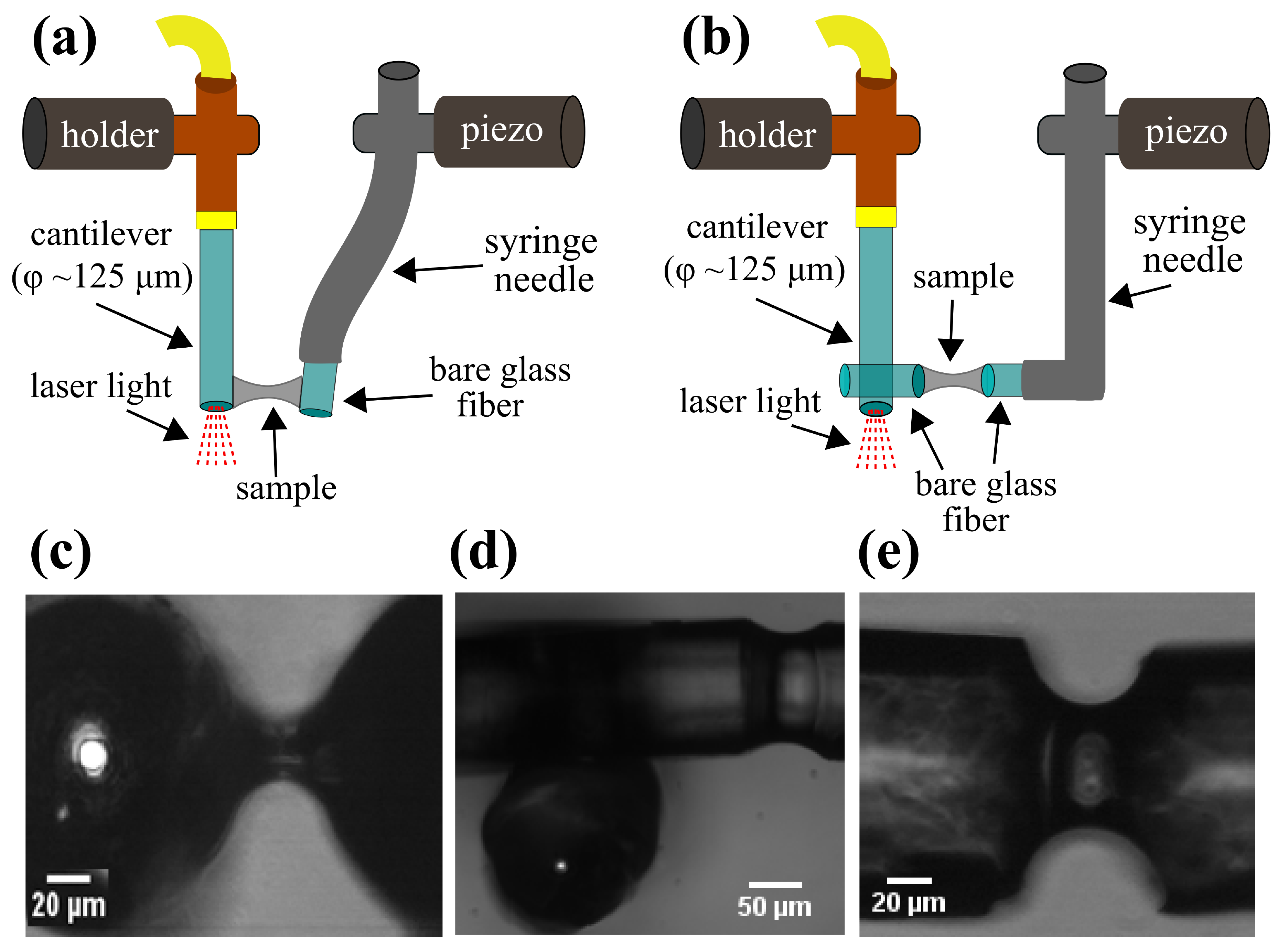}
\caption{
{\bf Schematics of different surface geometries.} {\bf (a)} In the simplest configuration of the MER, the sample is held between two cylindrical surfaces. While such experiments are easy to implement, the curved boundary conditions are not ideal. {\bf (b)} The issue of curved boundaries may be overcome by attaching very short pieces of cleanly cut optical fibers to the tips of the cantilever (diameter $\phi$) and the rigid support. This gives circular surfaces with well defined diameters. The side of the cantilever is tracked as before using the laser spot emanating from the vertical fiber. {\bf (c), (d)} Images of the two arrangements shown in schematics (a) and (b) respectively. The bright spots are the laser light exiting the cantilever. {\bf (e)} An expanded 
view of a sample suspended between horizontally attached end-pieces as shown in (b).
 }
\label{surfaces}  
\end{center}
\end{figure}

Several rheological processes such as electrospinning (of reconstituted silk), extrusion (of polymer melts), formation of thin films, {\it etc.} involve the extensional flow of non-Newtonian fluids \cite{mckinley2002, koeppel2018}. The properties of these materials under extensional deformation can be significantly different from that observed in shear due to the effects of strong flow alignment, surface effects, {\it etc.} \cite{mckinley2002, koeppel2018}. We have previously shown that the MER can be used for such studies, with the added advantage of being able to work with nanoliter volumes of sample, and with the ability to simultaneously image the surface profile \cite{paul2017}. For micro-scale samples, gravitational effects become negligible in comparison with effects due to surface tension forces, allowing fluid filaments to be studied in a horizontal geometry \cite{paul2017}. 

Various rheology protocols such as oscillatory extensional strain and exponential strain have been demonstrated \cite{paul2017, girithesis}. In the case of oscillatory extensional strain, a short fluid strand is held  between two optical fibers of identical diameter, one short and rigid and the other long and flexible (a cantilever) as shown in figure~\ref{exten-rheo}(a). An oscillatory strain is applied using the rigid fiber which is attached to the piezoelectric transducer. Both fibers are tracked using a high-speed CCD camera at a frame rate of 400 Hz \cite{paul2017}. From the images, Lissajous figures (over 25 oscillation cycles) of the extensional stress vs strain for different measured strain amplitude ($\epsilon_0$) and angular frequencies ($\omega$) are obtained. A Pipkin diagram of such Lissajous plots for polydimethylsiloxane (PDMS) is shown in figure~\ref{exten-rheo}. 

\begin{figure}
	\centering
	\begin{subfigure}{}
		\includegraphics[width=0.8\textwidth]{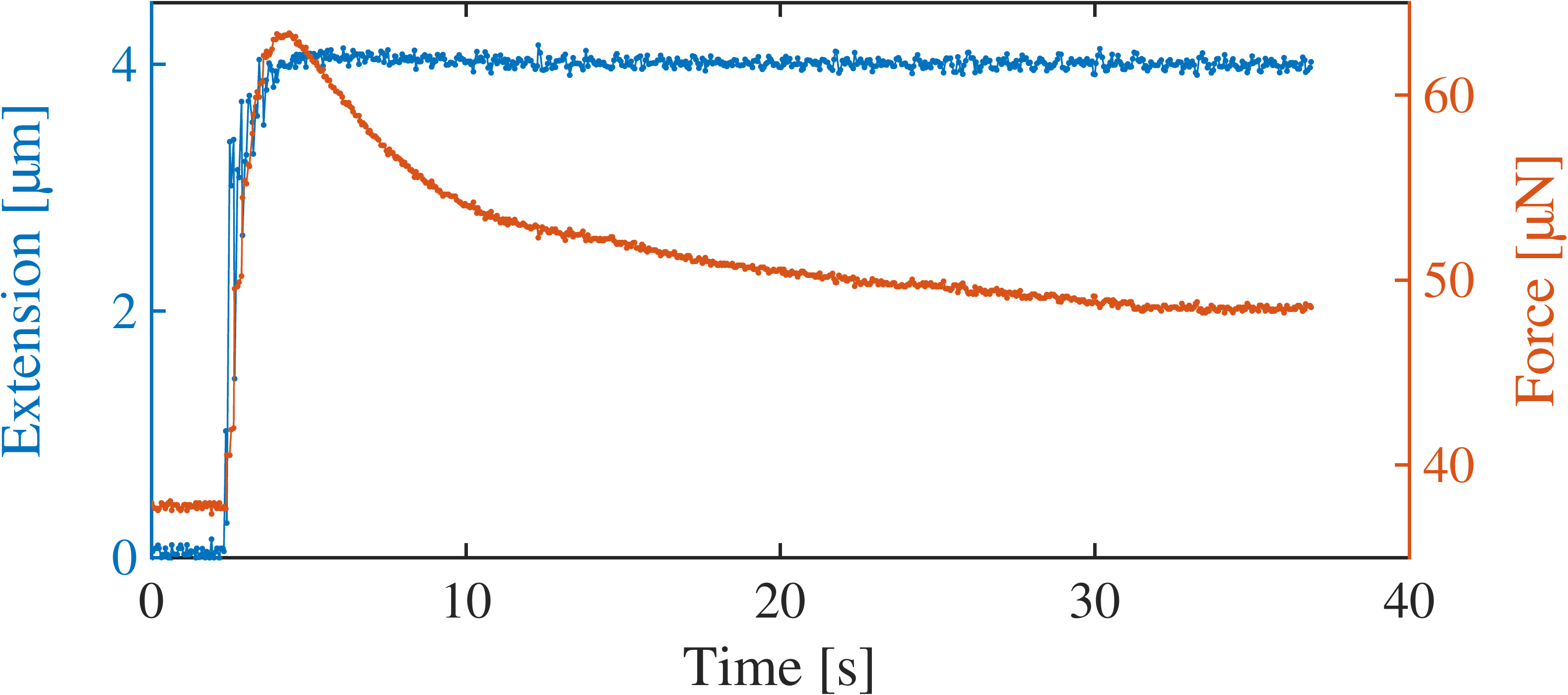}
	\end{subfigure}
	 \caption{A plot showing an imposed step-extension and the resultant force response for a filament of polydimethylsiloxane (PDMS).}
	\label{exp-strain}  
\end{figure}

High-molecular weight non-Newtonian fluids can also be subjected to an extensional flow by applying an exponential strain by moving the surface attached to the piezoelectric transducer \cite{mckinley2002}. The resulting stress is calculated by tracking the deflection of the cantilever using the PSD and by measuring the filament diameter at the mid-section of the strand. In the simplest configuration, the sample is held between two cylindrical surfaces as shown in figures~\ref{surfaces}(a), (c). Flat circular surfaces can also be obtained by cutting short pieces of the optical fiber and attaching them horizontally to the cantilever and to the rigid support as shown in figures~\ref{surfaces}(b), (d) and (e). A demonstration of an imposed step-extension, controlled using the feedback loop, and the resultant force response of a PDMS strand is shown  in figure~\ref{exp-strain}.

\subsection{\bf Probing mechanical responses of axons}

\begin{figure}[!h]
\begin{center}
\includegraphics[width=3.5in]{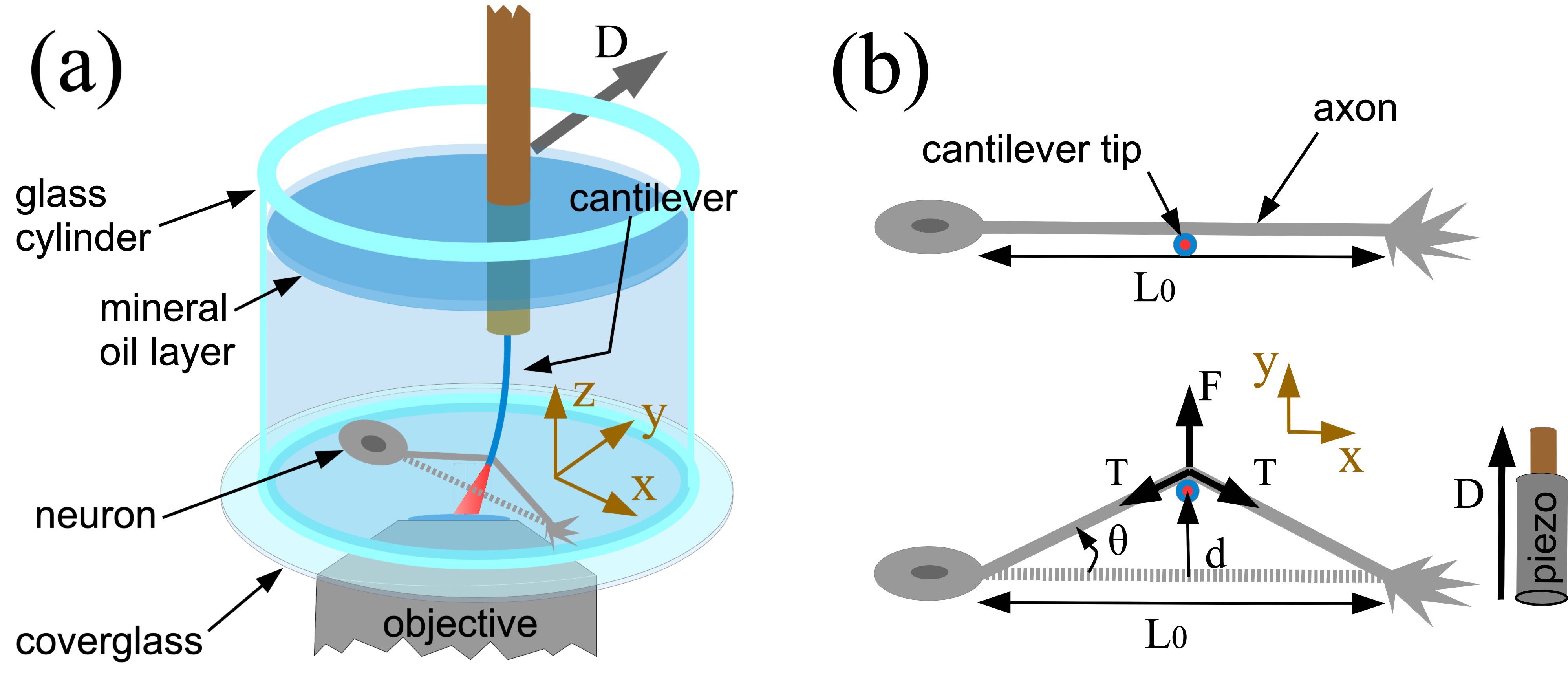}
\caption{ {\bf Schematic of axon stretching experiments.} {\bf (a)} Axons which are adherent at the two extremities - the cell body and the growing tip, are stretched by displacing their midpoints perpendicular to their length by placing the cantilever in mechanical contact (also see figure~\ref{setup-fluorescence}(c) and Supplementary Material Video-2). {\bf (b)} The stretching protocol and the parameters used for calculation of the axonal strain and the tension $T$ from the initial length $L_0$ and the cantilever tip displacement $d$ may be found in Ref. \cite{dubey2020-axon}. 
}
\label{geometry}  
\end{center}
\end{figure}

\begin{figure}[!h]
\begin{center}
\includegraphics[width=5in]{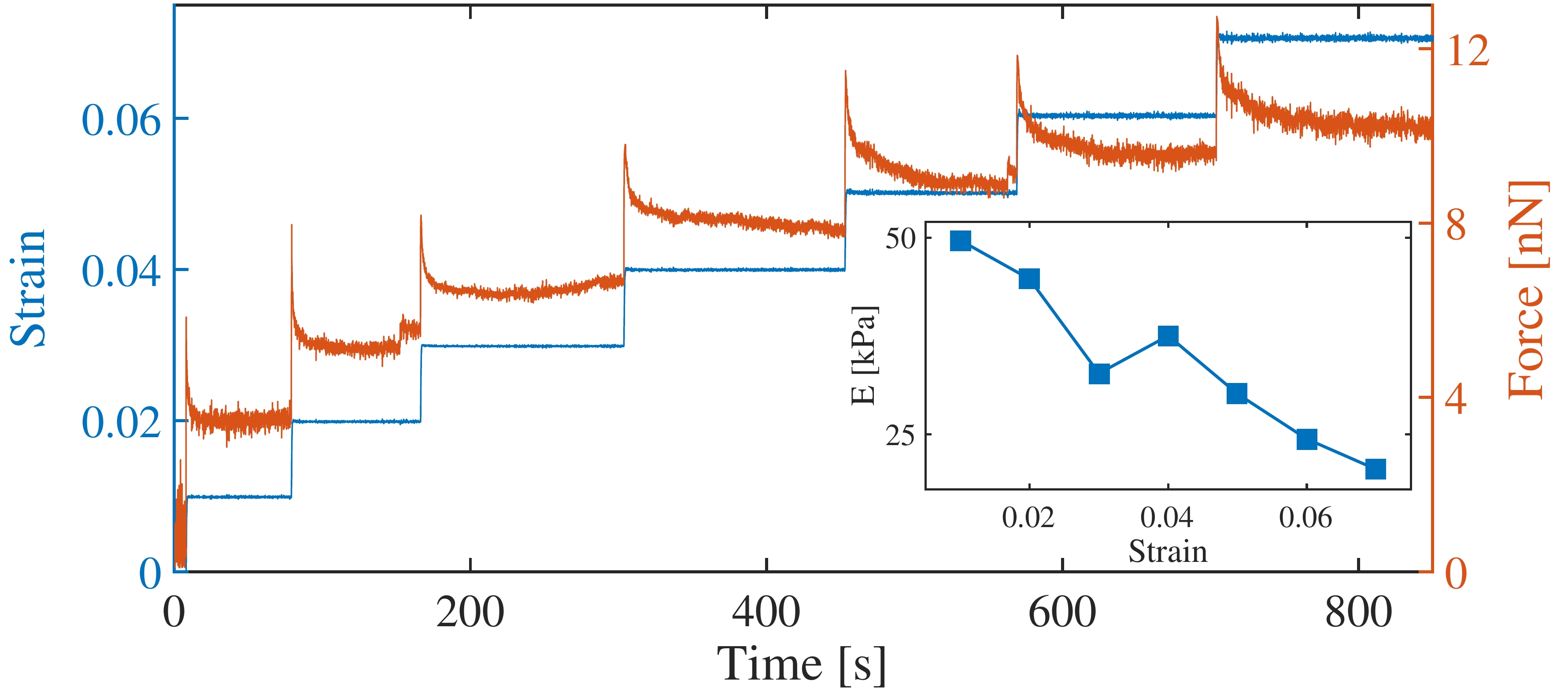}
\caption{ {\bf Nonlinear viscoelastic response of axons.} Axons were extended by applying sequential step strains with a wait time between steps during which the strain was held constant (Blue data). The resulting force response is shown in Orange. The viscoelastic nature of the axon shows up in the force relaxation after each step. The inset shows the Young's modulus $E$ calculated from the steady-state tension (see \cite{dubey2020-axon} for details), which exhibits a decreasing trend with increasing strain. Figure is reproduced from Ref. \cite{dubey2020-axon}. See Supplementary Material Video-3 for an animation.
}
\label{force-axon}  
\end{center}
\end{figure}

Axons are long tubular extensions generated by neuronal cells in order to transmit electrical signals within an organism. Due to their thin ($\sim$1 $\mu$m) and long (up to a meter in humans) geometry, they are subject to significant mechanical stresses, especially at the joints of limbs. Hence, understanding how axons respond to stretching is of importance, but challenging due to their microscopic dimensions and the need to maintain the neuronal cells in a living condition, attached to a surface \cite{bernal2007, mutalik2020}. Recently, we investigated the nonlinear mechanical properties of axons using the MER \cite{dubey2020-axon} via the scheme shown in figure~\ref{geometry}. In this study, living axons were stretched by applying sequential step strains with a wait time between steps during which the strain was held fixed via a feedback loop, and the ensuing force relaxation measured (figure~\ref{force-axon}). As axons are stretched by applying a lateral displacement at their midpoints, tension along the axon needs to be calculated. This is done by feeding the undeformed length $L_0$ to the computer and then computing the strain and tension using the angle of deformation $\theta$ which is obtained from the initial length and the displacement of the midpoint $d$ (figure~\ref{geometry}). The experiments revealed that axons undergo strain-softening with its zero-frequency Young's modulus decreasing with increasing steady-state strain as shown in figure~\ref{force-axon}. Experiments were also performed after perturbing specific biopolymer elements in the axons while subjecting them to a cyclic strain. These results helped identify the contributions arising from the various internal components of the axon to its mechanical response, and a
possible mechanism for strain-softening \cite{dubey2020-axon}.
\subsection{\bf Probing rheological properties of spider silk}

\begin{figure}[h]
\begin{center}
\includegraphics[width=3in]{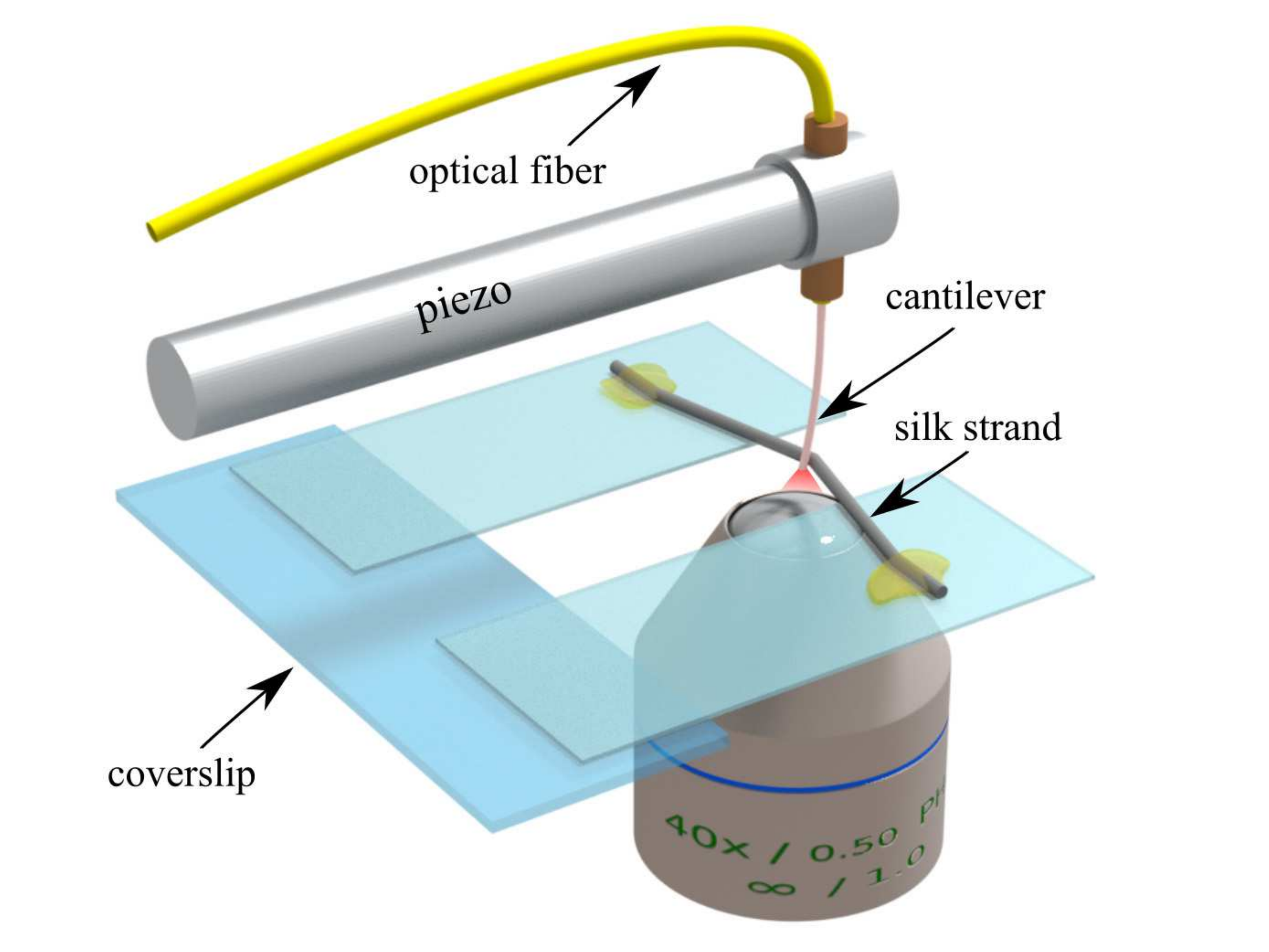}
\caption{{\bf Schematic showing a spider-silk strand stretched using the MER.} Silk fibers are transferred to a frame consisting of two rectangular coverslips with a $0.5$ mm gap, attached to a glass slide. The two ends of the silk strand are glued using a small quantity of superglue. The cantilever is placed at the midpoint of the strand to stretch it by imposing a displacement perpendicular to the strand using various rheological protocols. In this geometry, tension along the strand has to be calculated from measured force by measuring the angle that the strand makes with its initial configuration. The angle is fed to the computer before a measurement starts, and the displacement of the tip of the cantilever is measured in real time. Image reproduced with permission from the Royal Society of Chemistry \cite{dubey2019-silk}. 
}
\label{silk-schematic}  
\end{center}
\end{figure}
\begin{figure}[!h]
\begin{center}
\includegraphics[width=5.0in]{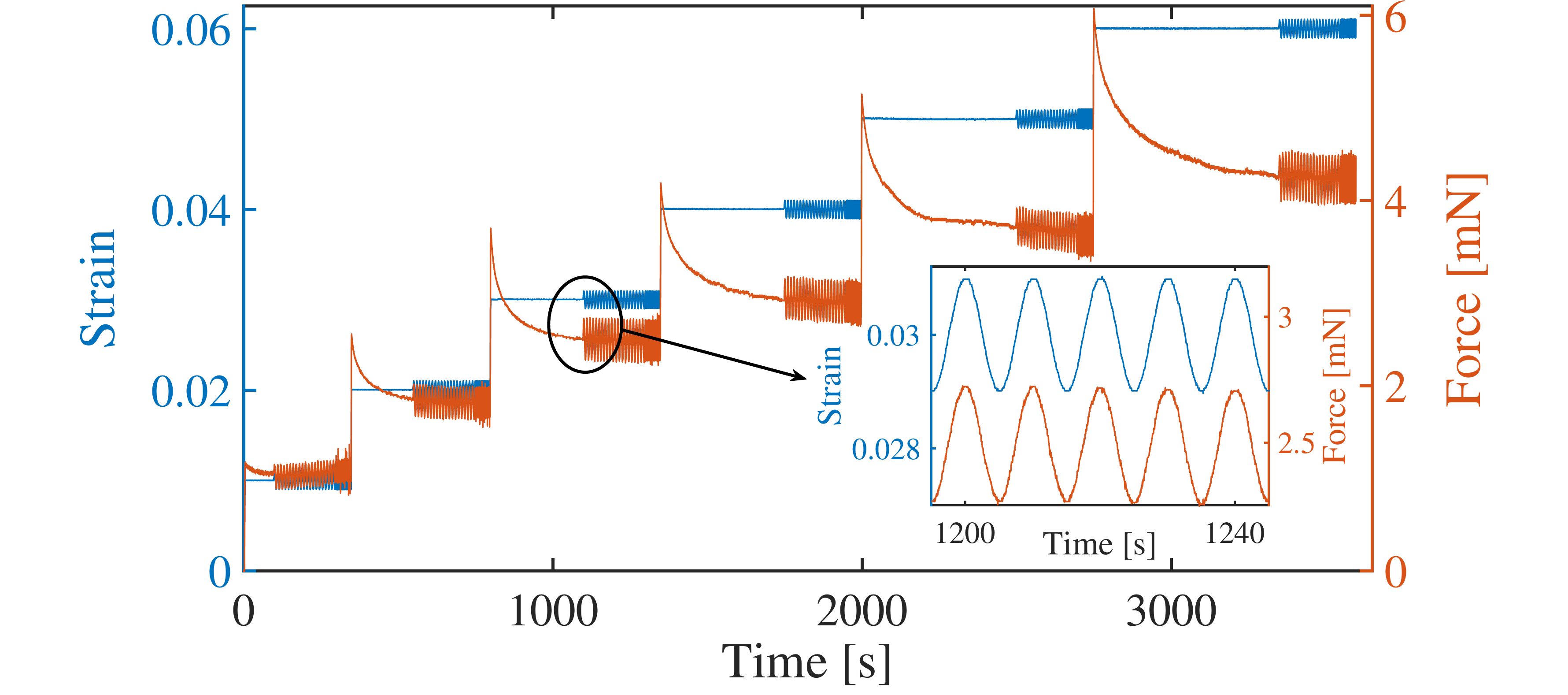}
\caption{{\bf Viscoelastic response of dragline silk.} Data obtained by implementing sequential step-strains along with superimposed, small-amplitude sinusoidal strain close to the steady-state force and the corresponding force response. Tension along the strand is calculated as mentioned in the caption to figure~\ref{silk-schematic} and detailed in Ref. \cite{dubey2019-silk}. The viscous component may be obtained from the phase-difference between the strain and force (or stress) oscillations, but for silk, no detectable phase-difference was observed. The inset shows an expanded view of the superimposed oscillations and corresponding force response.} 
\label{silk-force}  
\end{center}
\end{figure}

\begin{figure}[!h]
\begin{center}
\includegraphics[width=4in]{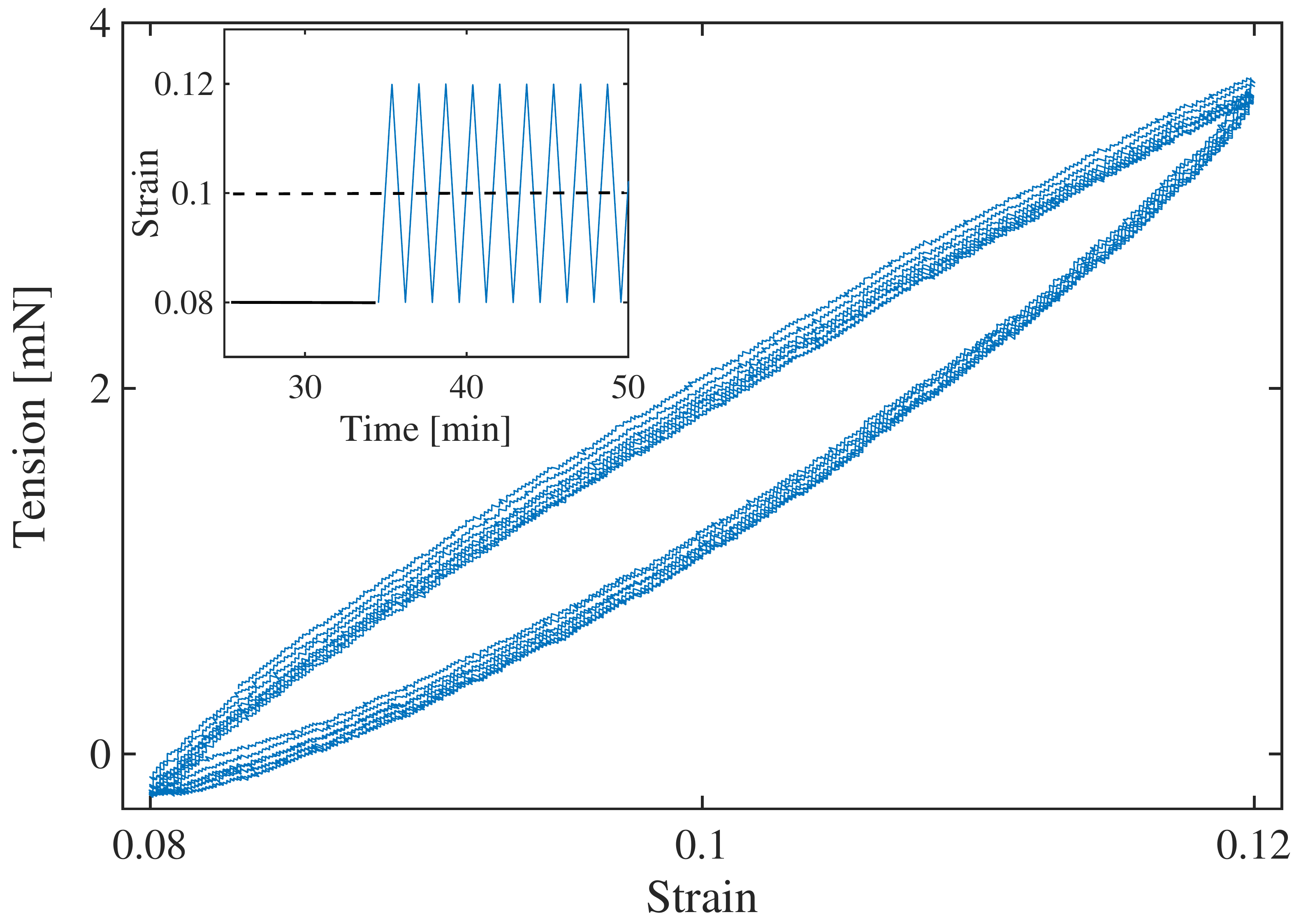}
\caption{{\bf Cyclic ramp-protocol.} Cyclic ramp applied on dragline silk with a pre-strain of 8\%, ramp amplitude of 4\% and ramp frequency of 0.01 Hz (see Supplementary Material Video-4). The hysteresis loop for tension vs strain is also shown. With this protocol, the energy dissipated per unit volume can be calculated from the area enclosed by the corresponding tensile stress-strain curve. 
}
\label{ramp}  
\end{center}
\end{figure}

Spider silk exhibits remarkable mechanical properties such as high tensile strength and a large yield strain and the origins of these responses is yet to be fully understood \cite{dubey2019-silk, yarger2018, harmer2011}. Most studies on silk are performed by applying large extensions upto yield point or large cyclic extensions \cite{yarger2018, harmer2011}. We have performed rheological measurements on silk using the MER in a variety of modes (figure~\ref{silk-schematic}). First, we probed the steady state moduli as a function of pre-strain by applying sequential step-strains and then superimposed small-amplitude oscillations as the strand approached a steady state stress (figure~\ref{silk-force}). Cyclic strains of different frequencies were also employed to study hysteresis loops as shown in figure~\ref{ramp}. A method was also developed to extract frequency response of the storage and loss moduli from step-strain data, further extending the capabilities of the MER \cite{dubey2019-silk}. 

Using these techniques, we observed that silk exhibits an initial strain-softening followed by a stiffening response \cite{dubey2019-silk}. The MER technique permits measurement of the time for stress relaxation and this shows an interesting behavior wherein two characteristic times increase monotonically with strain \cite{dubey2019-silk}. These experiments place additional constraints on the various rheological models describing the mechanics of silk \cite{tommasi2010}.

\subsection{\bf Probing nonequilibrium properties of bacterial baths}

\begin{figure}[!h]
\centering
\begin{subfigure}{}
	\includegraphics[width=1.5in]{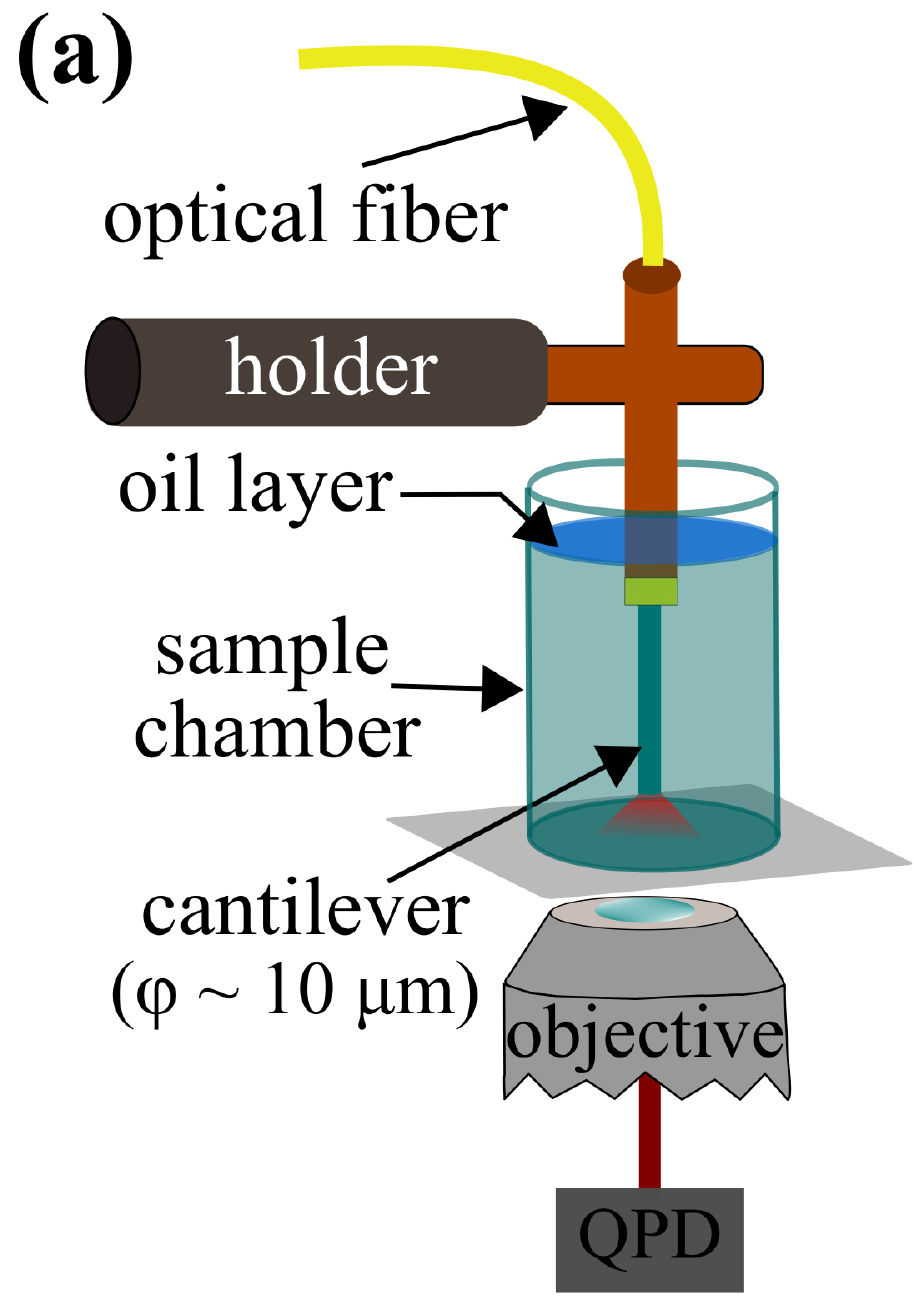}
	\end{subfigure}
\begin{subfigure}{}
	\includegraphics[width=3.5in]{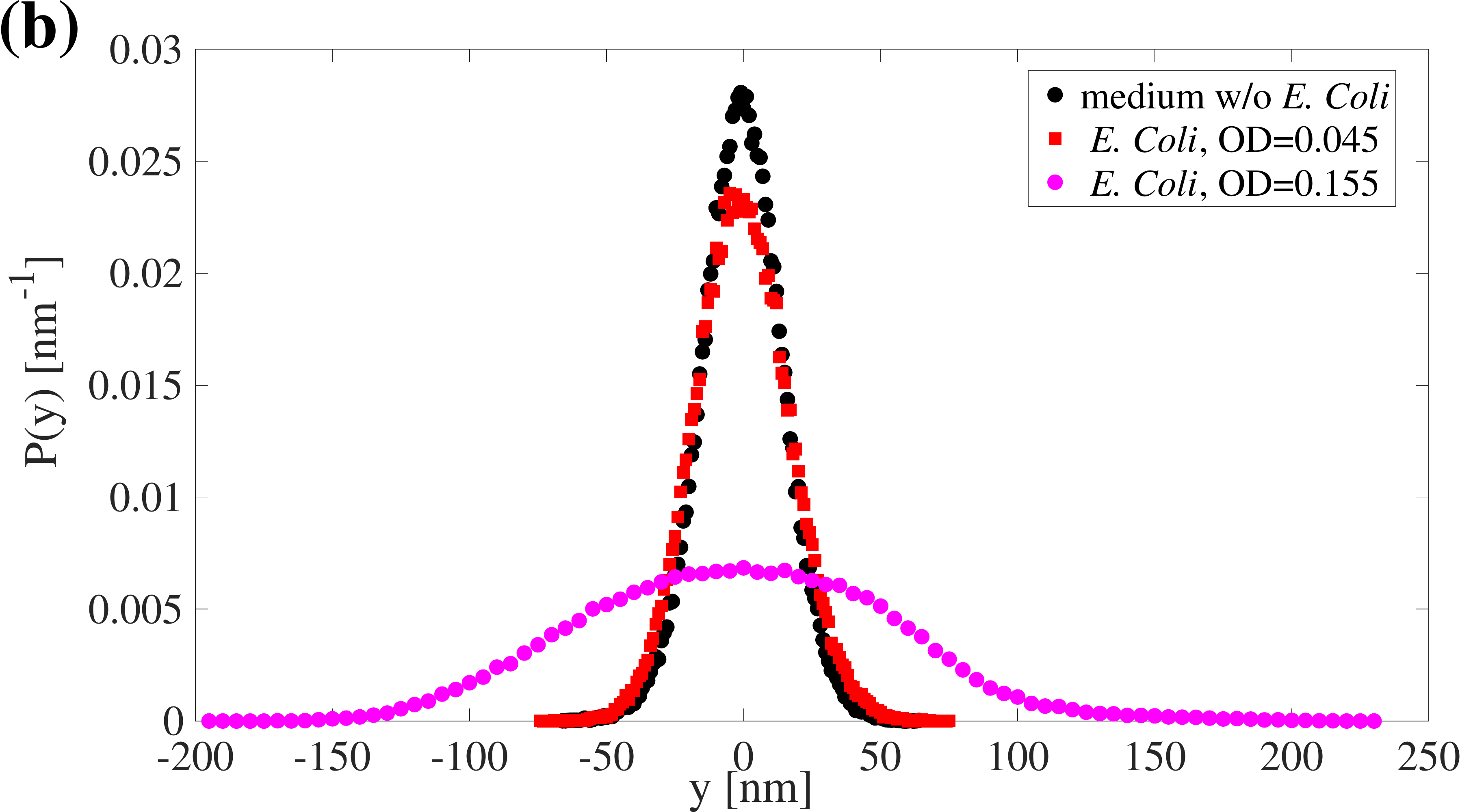}
	\end{subfigure}
\caption{{\bf Investigation of nonequilibrium properties of bacterial baths.} {\bf (a)} Schematic for investigation of the nature of bacterial baths. An etched optical fiber is suspended vertically in a bath containing {\it E. coli} bacteria. The fluctuation of the tip of the cantilever (diameter $\phi$) is recorded via a QPD. {\bf (b)} Probability distribution of the $y$-component of the displacement of the tip of the cantilever for bacterial baths with optical densities (OD) equal to 0.045 and 0.155. A measurement performed using a medium without bacteria is also shown for comparison. The position data has been detrended and filtered using a Savitzky-Golay filter prior to plotting.}
\label{bacteria-exp}  
\end{figure}

One of the simplest applications of an etched optical fiber, when used as a force-sensing cantilever, is to investigate the active or nonequilibrium nature of bacterial baths. In such studies, a culture of motile bacteria is maintained in a steady state and the dynamics of this active suspension is probed by monitoring the fluctuations of the tip of the cantilever. For this, an etched cantilever with typical dimensions of length 20 mm and diameter $\phi= $10 $\mu$m is immersed in the bacterial bath of interest (figure~\ref{bacteria-exp}(a)). The bath consists of the desired strain of {\it E. coli} suspended at a known density (measured using the OD$600$ method \cite{mol}) in a medium containing nutrients essential for maintaining the culture. This medium has a density close to that of water and behaves as a Newtonian fluid. A drop of mineral oil is then added to the surface of the medium to prevent convection caused by evaporative cooling of the top layer. In this application a QPD is preferred over a PSD because fluctuations are expected to be small and the QPD gives a spatial resolution $\sim4$ nm. The X and Y time series of the cantilever fluctuations are recorded using the QPD at a rate of 10 kHz. A plot of the probability distribution of the $y$-component of the displacement of the tip of the cantilever for {\it E. coli} at two different optical densities (OD) are shown in figure~\ref{bacteria-exp}(b). The probability distribution for the medium without bacteria is also shown for comparison. It can be seen that the width of the distributions increase progressively with the density of bacteria due to their active dynamics. {\it E. coli} are known to exert forces of the order of a piconewton and hence these experiments demonstrate the wide force sensitivity of the MER. 

\section{Conclusions}
We have demonstrated the usefulness of the Micro-Extensional Rheometer by performing a wide variety of rheological experiments on soft and living materials such as polymer melts, axons of neuronal cells, spider-silk fibers and bacterial suspensions. The main advantage of this instrument is its ability to apply a controlled tensile strain on microscopic samples such as living cells or by using samples available in volumes as low as a nano-liter. Stretching the sample in the image plane of a microscope has the advantage that the sample deformation and microscopic structural 
changes occuring during deformation can be simultaneously investigated. As the setup can be mounted on any commercial inverted microscope, the imaging schemes can easily be extended to include phase-contrast, birefringence, fluorescence and other sophisticated imaging techniques. The device offers a wide force range, from piconewtons to millinewtons, which is tunable by choosing fibers of a suitable length and diameter. Apart from the versatility in application which has been demonstrated here, the setup is also attractive from the perspective of cost, as it can be constructed using inexpensive lasers and optical fibers, especially in the Infra-Red range, and can operate with simple CCD cameras, linear actuators, and microscopes as well. Thus, the MER is ideally suited for the investigation of complex fluids, soft matter, and biological samples and awaits other future applications.

\subsection*{\bf Acknowledgements}

We thank the staff of the Raman Research Institute's Mechanical Engineering and Electronics Engineering groups for technical support.  We acknowledge
A. Lele,  A. Nisal, P. Szabo, S. Pinge, V. Singh,  T. Sridhar, and Y. Hatwalne for technical support and discussions, M. Arsalan for the silk-stretching schematic and animation, and S. Majumdar for a critical reading of the manuscript. CK acknowledges the Science and Engineering Research Board, Government of India for financial support via grant SB/S3/CE/071/2013. PP acknowledges financial support via the Ramanujan Fellowship, Department of Science and Technology, Government of India and grant BT/PR13244/GBD/27/245/2009 of the Department of Biotechnology, Government of India. 

\section*{References}
\bibliography{Ref-MER}
\include{./supply_v9_PP_CK}

\end{document}

%% file: supply_v9_PP_CK.tex
\renewcommand{\figurename}{Figure}
\renewcommand{\thefigure}{S\arabic{figure}}
\renewcommand{\thetable}{S\arabic{table}}

\graphicspath{{supply-fig/}}


%
%
%
%

\setcounter{figure}{0}    

\section*{SUPPLEMENTARY  MATERIALS:}

\section*{Testing of Micro-Extensional Rheometer protocols}
\begin{figure}[!h]
\centering
\includegraphics[width=3.0in]{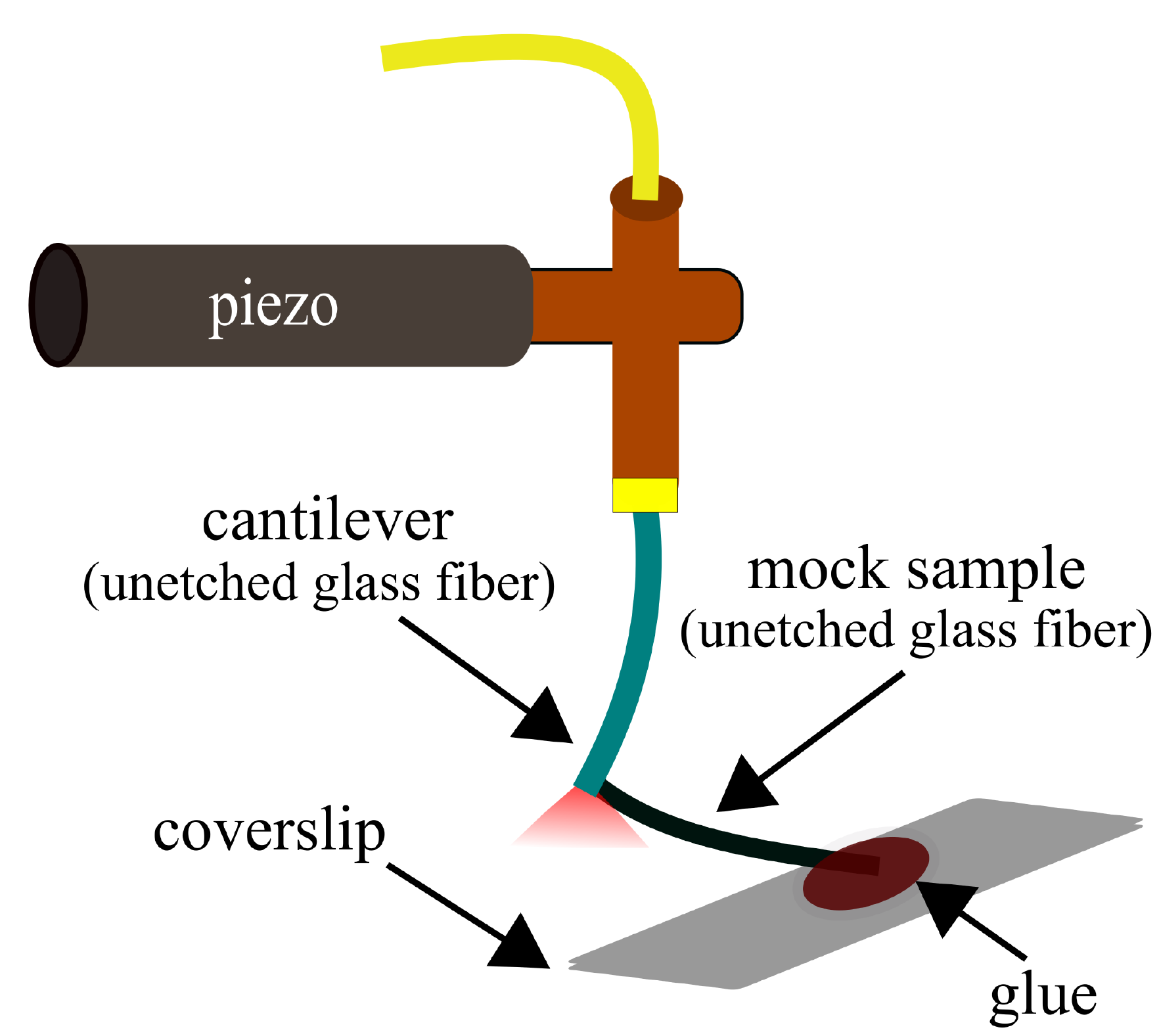}  
\caption{A schematic diagram showing the arrangement used to test the various operational protocols of the MER using a piece of optical fiber as a mock sample.
}
\label{mock-sample}  
\end{figure}

We have tested various operational protocols of the MER using a piece of optical fiber as a mock sample. The sample was attached horizontally to a surface and its tip displaced using the cantilever as shown in figure ~\ref{mock-sample}. Both the cantilever and the horizontally attached piece of fiber were unetched and of similar length. This arrangement was used to test the performance of the feedback loop and the operational protocols described in the main text.

\section*{Sequential step displacements}

\begin{figure}[!h]
\centering
\includegraphics[width=5.0in]{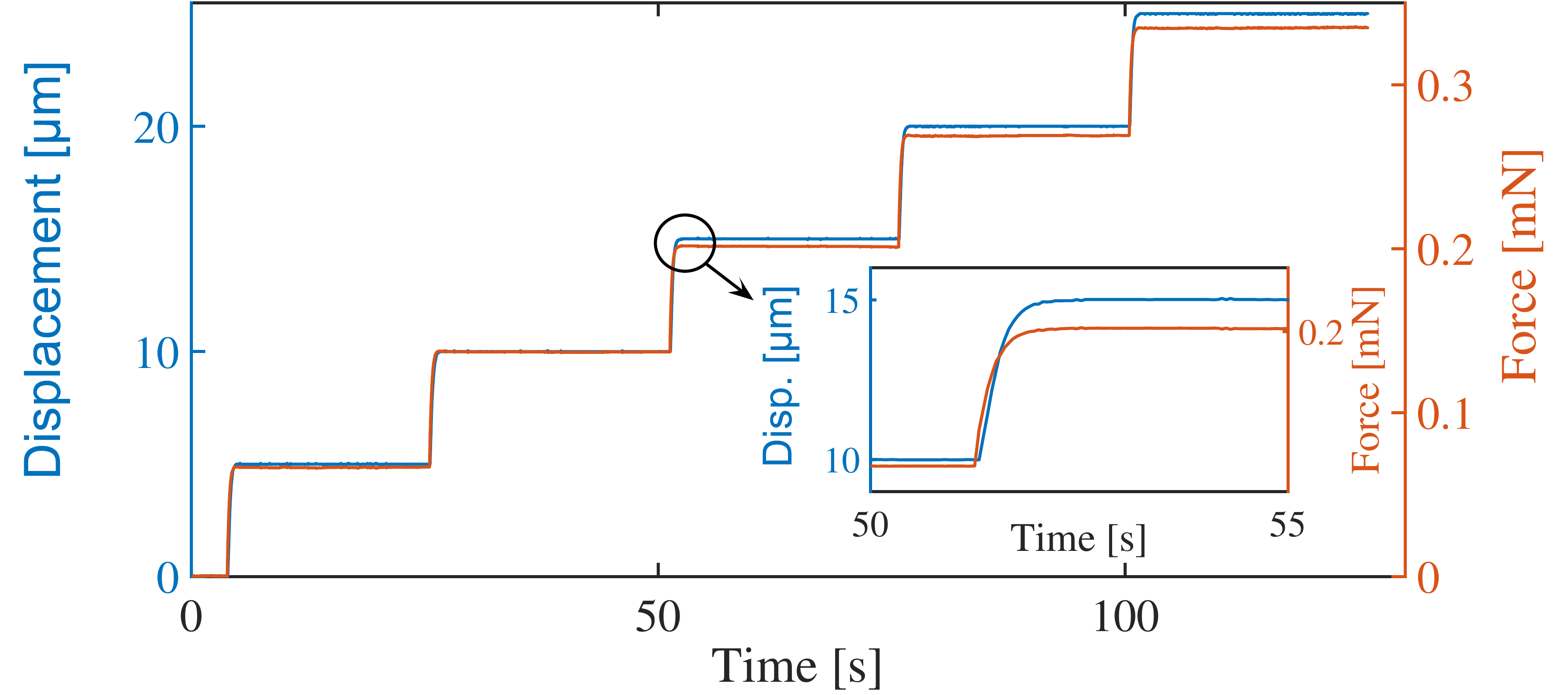}  
\caption{Force response of the mock sample obtained by applying sequential step displacements of 5 $\upmu$m with a wait time between steps. The displacement is controlled via a feedback loop as detailed in the main text. The inset shows an expanded view of a step. Note that there is no overshoot of displacement and the rise time is $\sim$1 s for a 5 $\upmu$m step.
}
\label{}  
\end{figure}
\newpage
\section*{Step strain with superimposed sinusoidal oscillations}

\begin{figure}[!h]
\centering
\begin{subfigure}{}
	\includegraphics[width=2.9in]{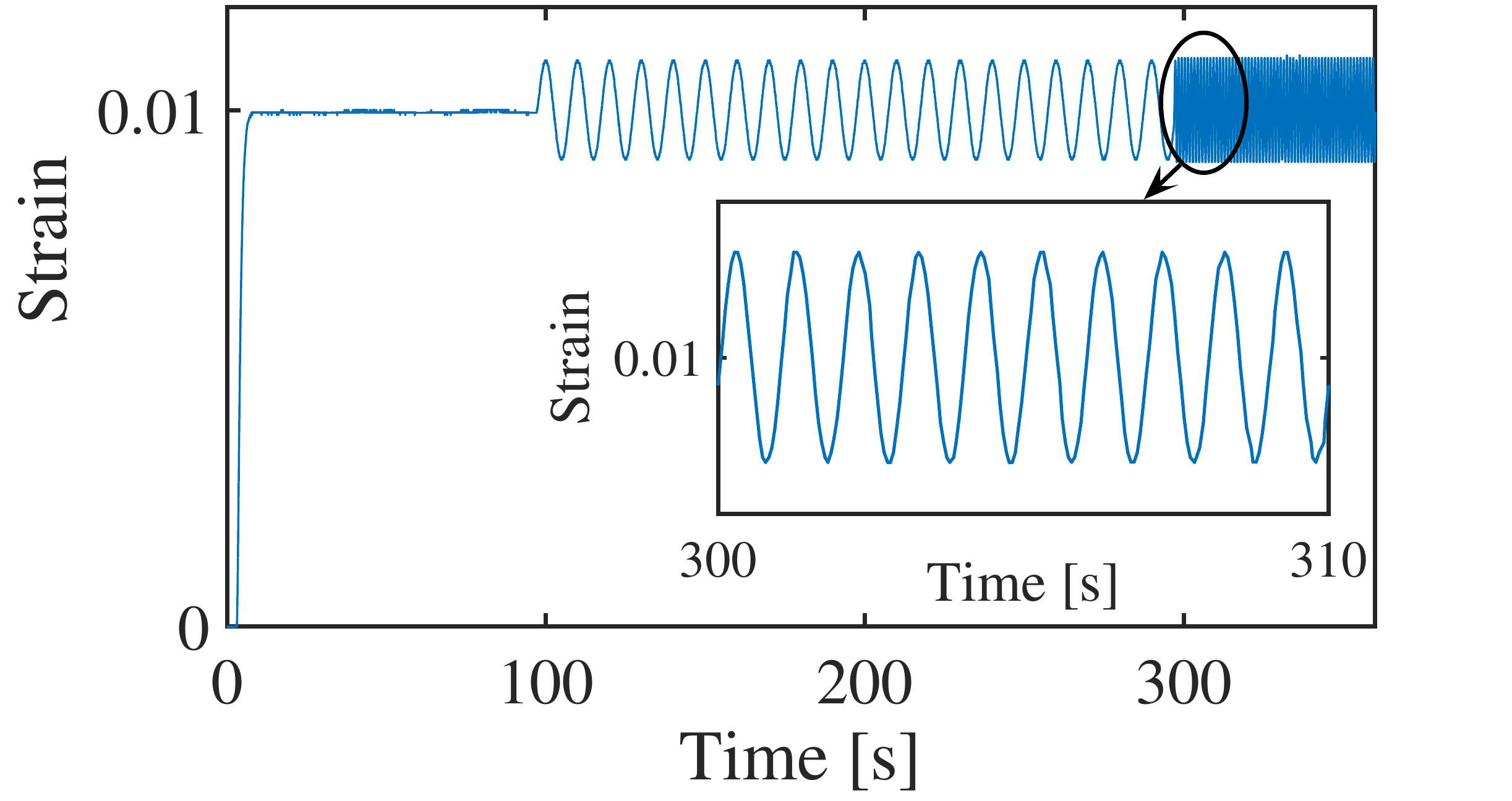}
	\end{subfigure}
\begin{subfigure}{}
	\includegraphics[width=2.9in]{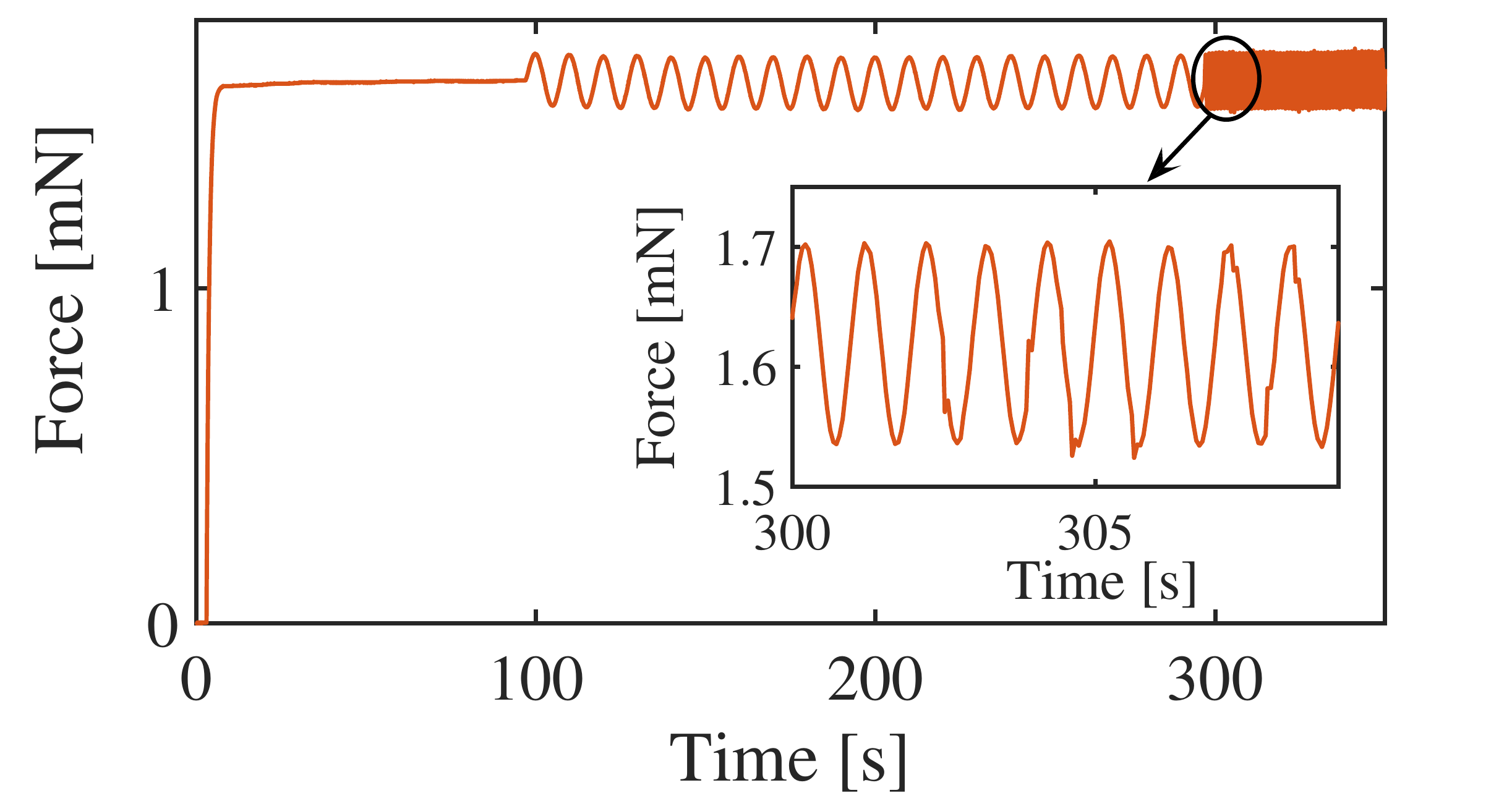}
	\end{subfigure}
\caption{In this test, sinusoidal oscillations are imposed after a step strain is applied to the mock sample. A wait time of $\sim$100 s permits the sample to reach a steady state before oscillations are imposed, first at 0.1 Hz and then at 1 Hz. The imposed strain as measured by the position-sensitive detector and the resulting force are shown. Insets show the expanded view for 1 Hz: Small deviations from a sine wave can be observed in the force waveform. These are further analysed via Fourier decomposition below. Note that the reference length used to compute strain is $200$ $\upmu$m, which is appropriate for a silk fiber.
}
\label{imposed-osc}  
\end{figure}

\begin{figure}[!h]

\begin{subfigure}{}
	\includegraphics[width=2.5in]{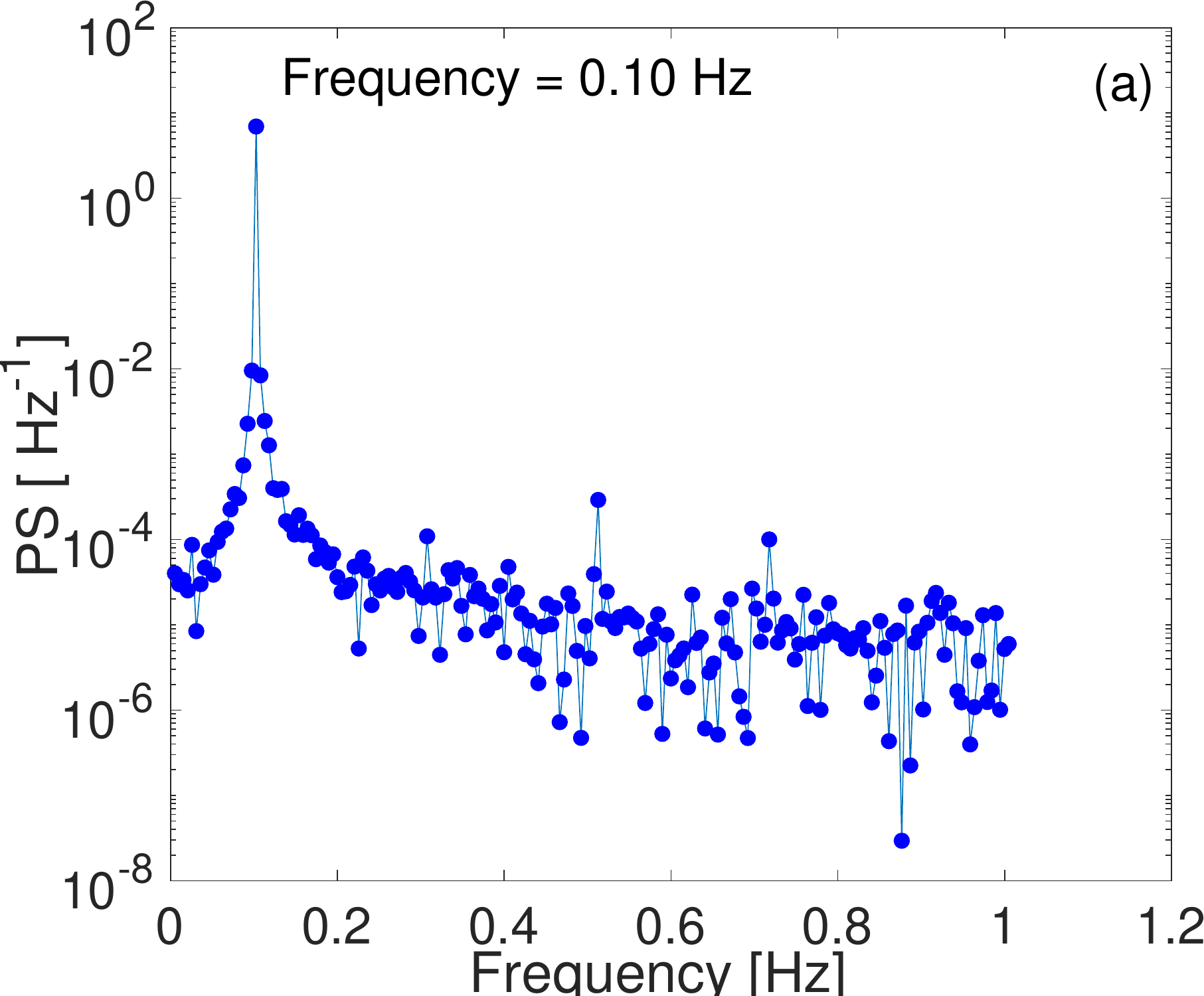}
	\end{subfigure}\hspace{1.0cm}
\begin{subfigure}{}
	\includegraphics[width=2.5in]{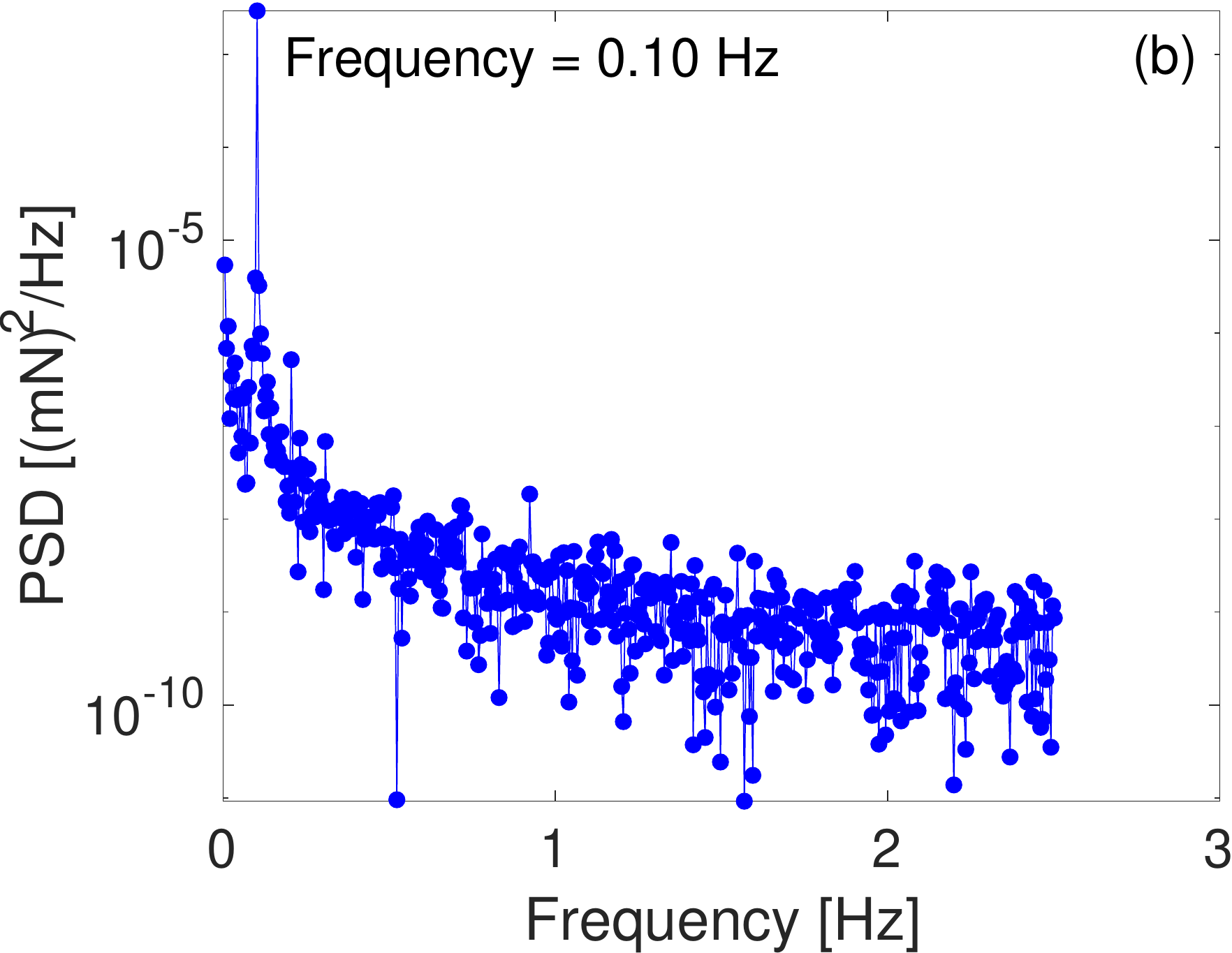}
	\end{subfigure}\hfill
	\begin{subfigure}{}
		\includegraphics[width=2.5in]{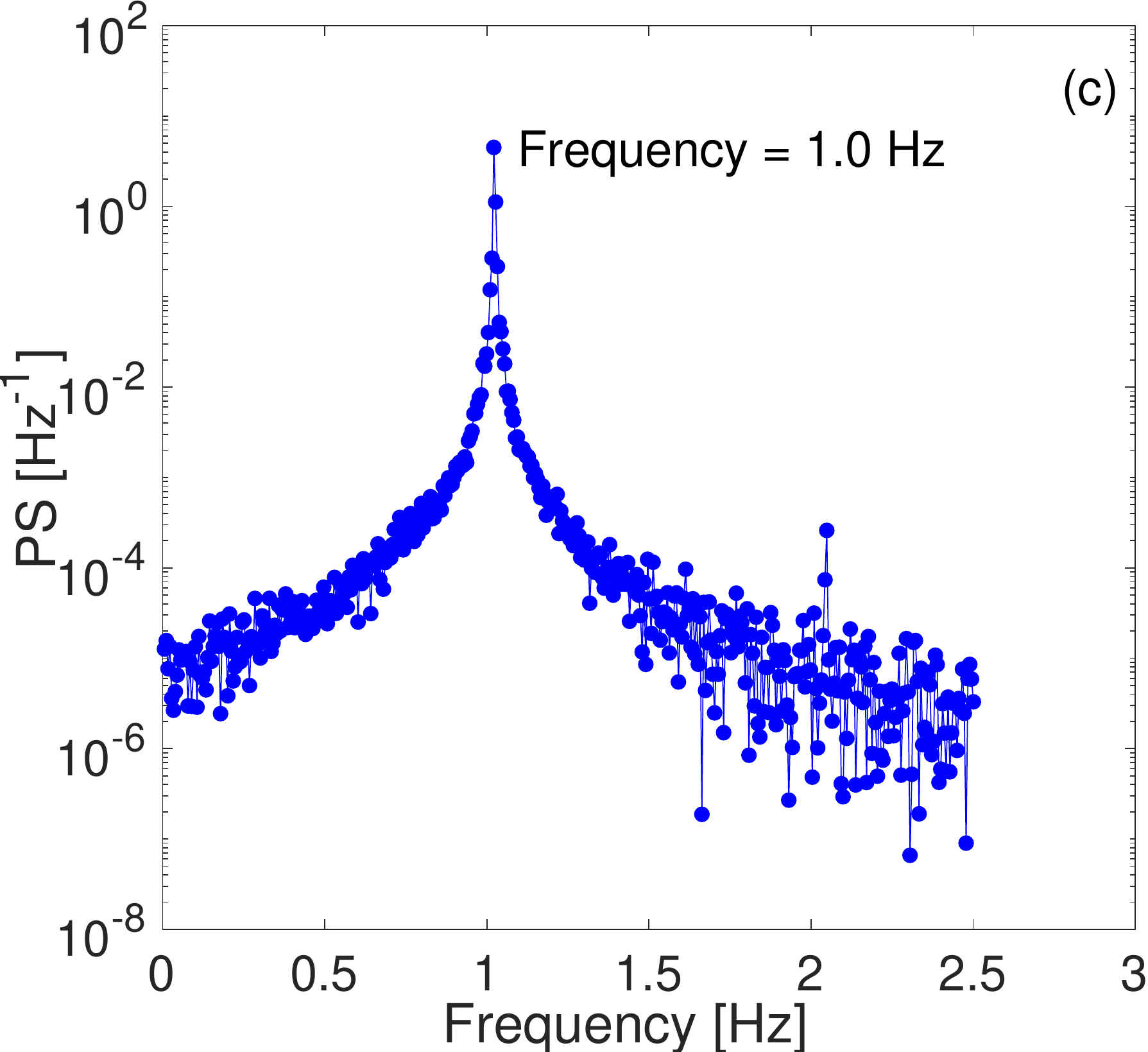}
		\end{subfigure}\hspace{1.0cm}
\begin{subfigure}{}
		\includegraphics[width=2.5in]{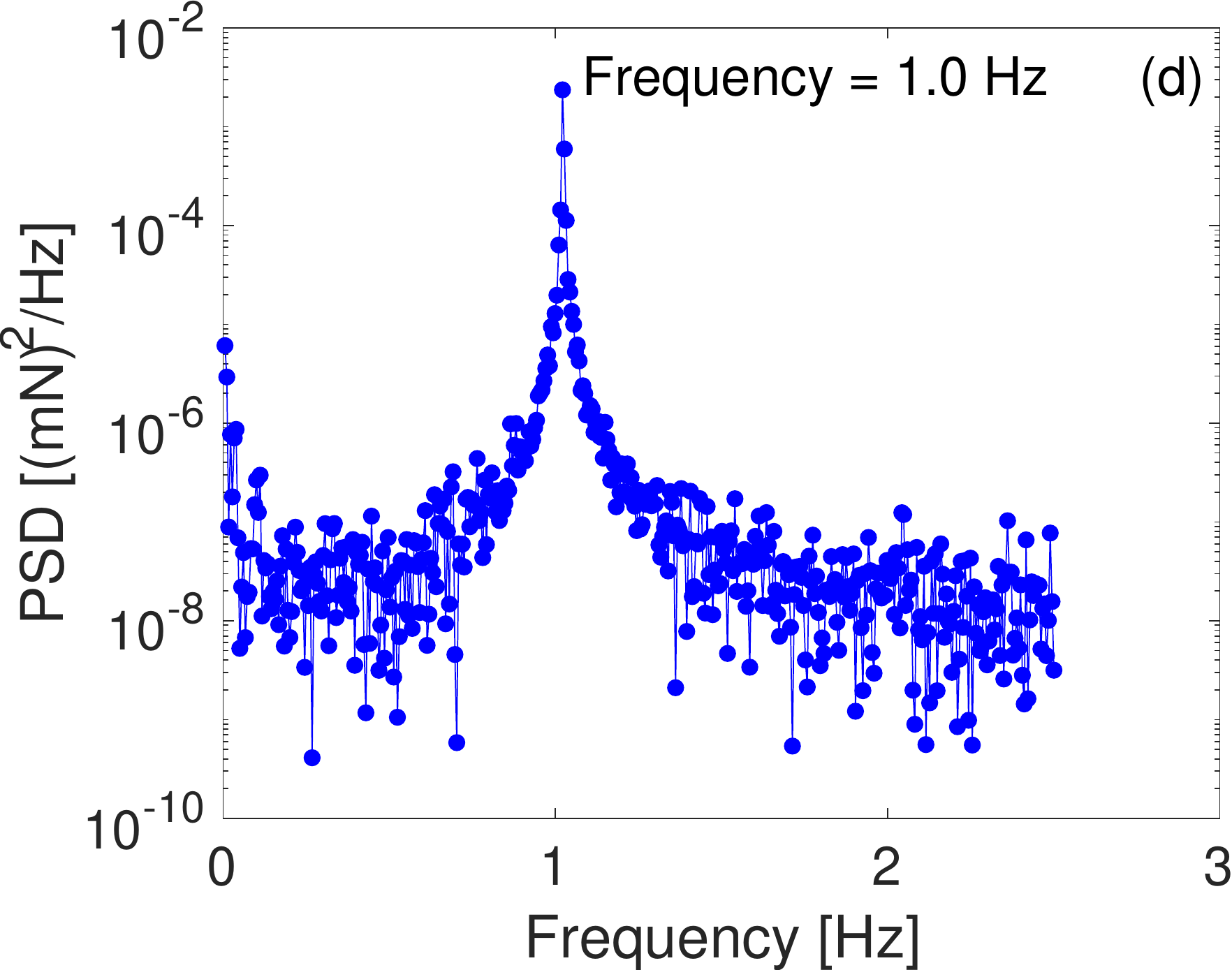}
		\end{subfigure}\hfill

	
\caption{Power spectra (PS) of the imposed oscillatory strain and the measured force waveforms obtained using the mock sample. The value of the peak frequency are indicated in each subfigure 
(a) PS of oscillatory strain at 0.1 Hz (b) PS of response force (imposed oscillatory strain of 0.1 Hz) (c) PS of oscillatory strain at 1 Hz (d) PS of response force  (imposed oscillatory strain of 1 Hz). The data shows that the MER can apply $\le$1 Hz sinusoidal oscillations without significant distortion in the force waveform.
}
\label{freq-osc}  
\end{figure}
\newpage
\section*{CAPTIONS FOR VIDEOS}

$\;$ \\
{\bf Video 1:} Video showing a sample of polydimethylsiloxane held between two cylindrical surfaces - one, a bare optical fiber that acts as the cantilever and the other a short piece of a bare optical fiber that acts as a rigid support. The sample is subjected to oscillatory extensional strain by moving the rigid surface which is attached to the piezoelectric transducer.\\ \\
{\bf Video 2:} Video showing an axon ($\simeq 1\; \upmu$m diameter) being stretched using the sequential step-strain protocol. The image is recorded using phase-contrast microscopy. A small fraction of the laser light exiting the optical core of the cantilever can be seen as a bright spot.\\ \\
{\bf Video 3:} An animation of the MER showing how a neuronal cell is stretched using the sequential step-strain protocol discussed in the main text. The inset shows a plot of the force response. \\ \\
{\bf Video 4:} Video micrograph of a strand of dragline silk ($\simeq 5\; \upmu$m diameter) being subjected to sinusoidal oscillatory strain. A unetched fiber is used in this case and the laser spot is visible.